\documentclass[sigconf, nonacm=true]{acmart}
\renewcommand\footnotetextcopyrightpermission[1]{} 
\usepackage{graphicx}
\usepackage{subfigure}
\usepackage{epstopdf}
\usepackage{multirow}
\AtBeginDocument{%
  \providecommand\BibTeX{{%
    \normalfont B\kern-0.5em{\scshape i\kern-0.25em b}\kern-0.8em\TeX}}}

\begin{document}

\title{Tracking Public Opinion in China through Various Stages of the COVID-19 Pandemic}

\author{Yuqi Gao}
\affiliation{%
  \institution{Software Institute\\Nanjing University}
}
\email{gaoyq@smail.nju.edu.cn}

\author{Hang Hua}
\affiliation{%
  \institution{
  Department of Software Engineering 
  \\ Peking University}
}
\email{huahang@pku.edu.cn}

\author{Jiebo Luo}
\affiliation{%
  \institution{Department of Computer Science University of Rochester}
}
\email{jluo@cs.rochester.edu}

\renewcommand{\shortauthors}{Tracking Public Opinion in China through Various Stages of the COVID-19 Pandemic}
\begin{abstract}
In recent months, COVID-19 has become a global pandemic and had a huge impact on the world.
People under different conditions have very different attitudes toward the epidemic.
Due to the real-time and large-scale nature of social media, we can continuously obtain a massive amount of public opinion information related to the epidemic from social media.
In particular, researchers  may ask questions such as "how is the public reacting to COVID-19 in China during different stages of the pandemic?", "what factors affect the public opinion orientation in China?", and so on. To answer such questions, we analyze the pandemic related public opinion information on Weibo, China's largest social media platform.
Specifically, we have first collected a large amount of COVID-19-related public opinion microblogs.
We then use a sentiment classifier to recognize and analyze different groups of users' opinions.
In the collected sentiment orientated microblogs, we try to track the public opinion through different stages of the COVID-19 pandemic.
Furthermore, we analyze more key factors that might have an impact on the public opinion of COVID-19 (e.g., users in different provinces or users with different education levels).
Empirical results show that the public opinions vary along with the key factors of COVID-19.
Furthermore, we analyze the public attitudes on different public-concerning topics, such as staying at home and quarantine.
In summary, we uncover interesting patterns of users and events as an insight into the world through the lens of a major crisis.
\end{abstract}


\maketitle

\section{Introduction}
The outbreak of COVID-19 is officially recognized as a pandemic by the World Health Organization(WHO) on March 11, 2020. The pandemic has made a huge impact on the world today. People can clearly feel the impact of the epidemic. In China, the COVID-19 epidemic has generated an outburst of public opinions in the Chinese Sina-Weibo. In this paper, we try to answer the question of how public opinion change with the development of COVID-19 pandemic in China and figure out what key factors may cause the change of public opinion. Since public sentiment is a good indicator of public opinion, so we disentangle these problems by analyzing the sentiment changes of the public on social media websites.

We divide the collected data into different groups according to two criteria: 
(1) geographical differences and (2) educational differences.
Since the first case infected by COVID-19 was identified in Wuhan in December 2019, multiple countries and regions reported infected individuals.
During different stages of the outbreak, people in different regions showed different sentiment orientations. 
Education background is another factor we are interested in. People with different education levels may have different opinions on certain social events about COVID-19. 

People's attitudes towards the Chinese and United States governments during the COVID-19 outbreak are also interesting, since the government issued policies that directly influence people's daily lives. Therefore, we analyze people's opinions towards the governments of China and the United States.

Our main contributions can be summarized as follows:
\begin{itemize}
        \item We collect large-scale data from Sina-Weibo and analyze public opinion on COVID-19 using textual information.

        \item We analyze and find different factors (e.g., education levels, regions, gender, epidemic trends) that affect the orientation of the public sentiment towards the Chinese and United States governments and social events in China.
        
        \item The extensive analyses show that our collected data are informative and the factors we analyzed have a significant impact on the public opinion.
\end{itemize}

\section{Related Works}
In recent years, due to the booming development
of online social networks, web information plays a significant role in shaping people's beliefs and opinions. With misinformation and disinformation,
such online
information can easily affect online social network users, in turn having tremendous effects
on the offline society. Therefore, public opinion analysis is important for monitoring and maintaining social stability.

Research studies on social media have pointed out how social media reflects \cite{j2} or affects \cite{o2011impact} the thoughts of different social groups. \citet{Badawy2018AnalyzingTD} analyze the digital traces of political manipulation related to the Russian interference of the 2016 US Presidential Election in terms of Twitter users' geo-location and their political ideology~\cite{b15}. Wang et al. compare the Twitter followers of the major US presidential candidates~\cite{j3, j6} and further infer the topic preferences of the followers~\cite{j4}. More closely related to this study, \cite{hutto2014vader, Lu2015VisualizingSM} explore the impact that disasters have on the underlying sentiment of social media streams. Our research draws knowledge from the body of research on characterizing the demographics of social media users, along the dimensions such as gender \cite{j6, rao2010classifying, Bergsma2013UsingCC}, age \cite{Nguyen2013HowOD, sloan2015tweets}, and social class~\cite{sloan2015tweets, agichtein2008finding}.

Sentiment analysis is a popular research direction in the field of social media. In this field, may natural language processing (NLP) technologies are employed to capture the public sentiment towards certain social events and analyze the causality of the public sentiment. The majority of past approaches employed traditional
machine learning methods such as logistic regression, SVM,
MLP, and so on trained on lexicon features and sentiment-specific
word embeddings (vector representations of words)~\cite{Maas2011LearningWV, Giahanou2016LikeIO}. Best performing models of this breed include \citet{Thongtan2019SentimentCU} which proposes training document embeddings using cosine similarity and achieves state-of-the-art on the IMDB dataset~\cite{Maas2011LearningWV}. \citet{Yin2015DynamicUM} use Distributional Correspondence Indexing (DCI) - a transfer learning method for cross-domain sentiment classification and achieve the first place on the Webis-CLS-10 dataset~\cite{Prettenhofer2011CrossLingualAU}. In our study, we collected 99,913 sentiment-labeled 
Weibo posts and 900,000 unlabeled Weibo posts. To make the samples more representative and improve the reliability of the analysis results, we bootstrap the sentiment labels using a sentiment classifier. Finally, we use a classifier to  predict the education background labels for users.

There are already some qualitative and quantitative analysis works related to social media information of COVID-19. \citet{yin2020covid} propose a multiple-information susceptible-discussing-immune model to understand the patterns of key information propagation on the  social networks.  \citet{Cinelli2020TheCS} address the diffusion of information about COVID-19 with massive data  on Twitter, Instagram and YouTube. The main difference between our work from these works is that we try to track the Chinese public opinion during different stages of the COVID-19 pandemic and analyze some key factors (e.g., 
education levels, gender, region, 
epidemic trends) that might have an impact on the public opinion of COVID-19.

\section{Data and Methodology}
\subsection{Data collection}
We collect a large-scale Sina-Weibo corpus from two sources. First, we use the dataset provided by the Data Challenge of The 26th China Conference on Information Retrieval (CCIR 2020).\footnote{\url{https://www.datafountain.cn/competitions/423/datasets}} as the seed data for classifier training. Second, we crawled microblogs on Sina-Weibo with COVID-19-related keywords.
After we obtained the COVID-19-related microblogs, we further collected the corresponding user information from Sina-Weibo and the number of crawled user profiles is 710,073.
The first data source covers the microblogs from Jan. 1 to Feb. 18 and the second data source covers the microblogs from Feb. 19 to Apr. 15.
According to the epidemic trend (the number of newly infected is descending in China and the number of newly infected is increasing outside China), data from Jan. 1 to Feb. 18 is marked as stage 1 and data from Feb. 19 to Apr. 15 is marked with as stage 2.

\subsection{Classifier}
Our collected dataset contains 999,13 Weibo microblogs with manually labelled sentiment polarity (positive, negative or neutral). We use these data to train a sentiment classifier. Specifically, we use the Fasttext \footnote{\url{https://fasttext.cc/}} framework to implement the classifier. We use 30\% of the labelled data to validate the classifier and its precision is 68\%. Based on our experience, this is on par with the performance of VADER~\cite{hutto2014vader} on tweets. We then use the classifier to predict the sentiment polarity for the remaining unlabeled data.

For the topics of concern, the corresponding keywords and similar expressions are used to filter the related microblogs.

\subsection{Sentiment Analysis}
Using the timelines of the COVID-19 pandemic summarized by
Wikipedia\footnote{\url{https://en.wikipedia.org/wiki/Timeline_of_the_COVID-19_pandemic}},
Ding Xiang Yi Sheng\footnote{\url{https://dxy.com/}} and China Daily\footnote{\url{http://www.chinadaily.com.cn/}},
we are able to identify key events during different stages of the pandemic.
These key events and the Weibo data with sentiment label enable us to track the public opinion with the sentiment polarity.
In order to provide an intuitive measure of the public opinion, we define a Sentiment Index as follows:
\begin{equation}
Sentiment Index = \frac{(Positive - Negative)}{(Positive + Negative)}
\label{eqn:sentiment index}
\end{equation}
where $Positive$ or $Negative$ represents the number of positive or negative microblogs.
The Sentiment Index varies in the range of $(-1, 1)$, where $1$ represents pure positive and $-1$ represents pure negative (ignoring neutral microblogs).
We build the index to capture the overall trend of the public sentiment.
\section{Empirical Results}
\subsection{Volume}
\begin{figure}[t]
  \includegraphics[width=0.45\textwidth]{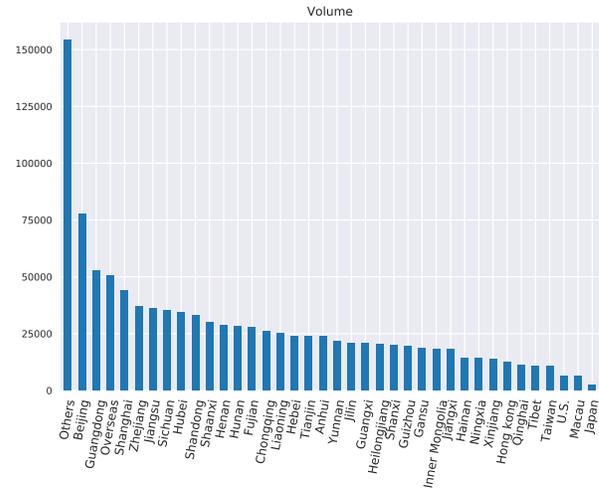}
  \vspace{-0.3cm}
  \caption{Where are the microblogs on the pandemic from?}
  \label{figure:region}
  \vspace{-0.3cm}
\end{figure}

In this subsection, we mainly discuss this question:
Who is discussing COVID-19 on the Internet considering the geographical distribution?
Based on the geographical information provided by the users, Figure \ref{figure:region} showss the number of uploaded microblogs from different regions.
`Other' refers to users who mark there locations with the label `Other'.
Because the U.S. is the world's only current superpower, Japan is near China and issued quite different policies compared with the U.S., we list these two representative countries separately.
It should be noted that `Overseas' refers to overseas users except the users whose profiles are labelled with `U.S.' or `Japan'. In other words, `Overseas' refers to all countries other than China, the U.S., and Japan. 

As shown in Figure \ref{figure:region}, a preliminary observation is that the number of microblogs from the regions with higher GDP per capita is more than the lower GDP regions considering the administrative divisions of China.
For example, Beijing and Shanghai discuss the pandemic even more than the most intensely hit areas by the pandemic, such as Hubei.

\subsection{Overall Sentiment}
This subsection intends to answer two questions:
How does the public sentiment vary with different stages of COVID-19 pandemic?
What are the public opinions of different groups of users?
\vspace{-0.2cm}
\subsubsection{Public opinion on different stages}
\label{Public opinion on different stages}
\begin{figure}[t]
  \includegraphics[width=0.42\textwidth]{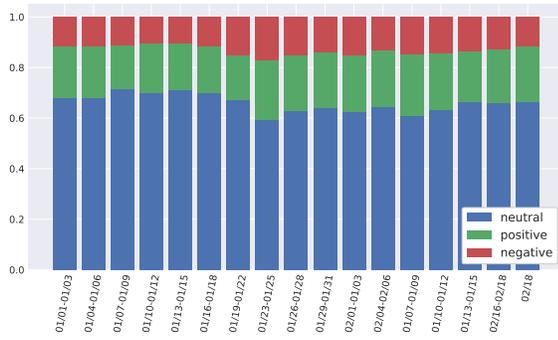}
  \vspace{-0.3cm}
  \caption{Sentiment proportion of stage 1}
  \vspace{-0.3cm}
  \label{figure:stage1_sentiment}
\end{figure}

\begin{figure}[t]
  \includegraphics[width=0.42\textwidth]{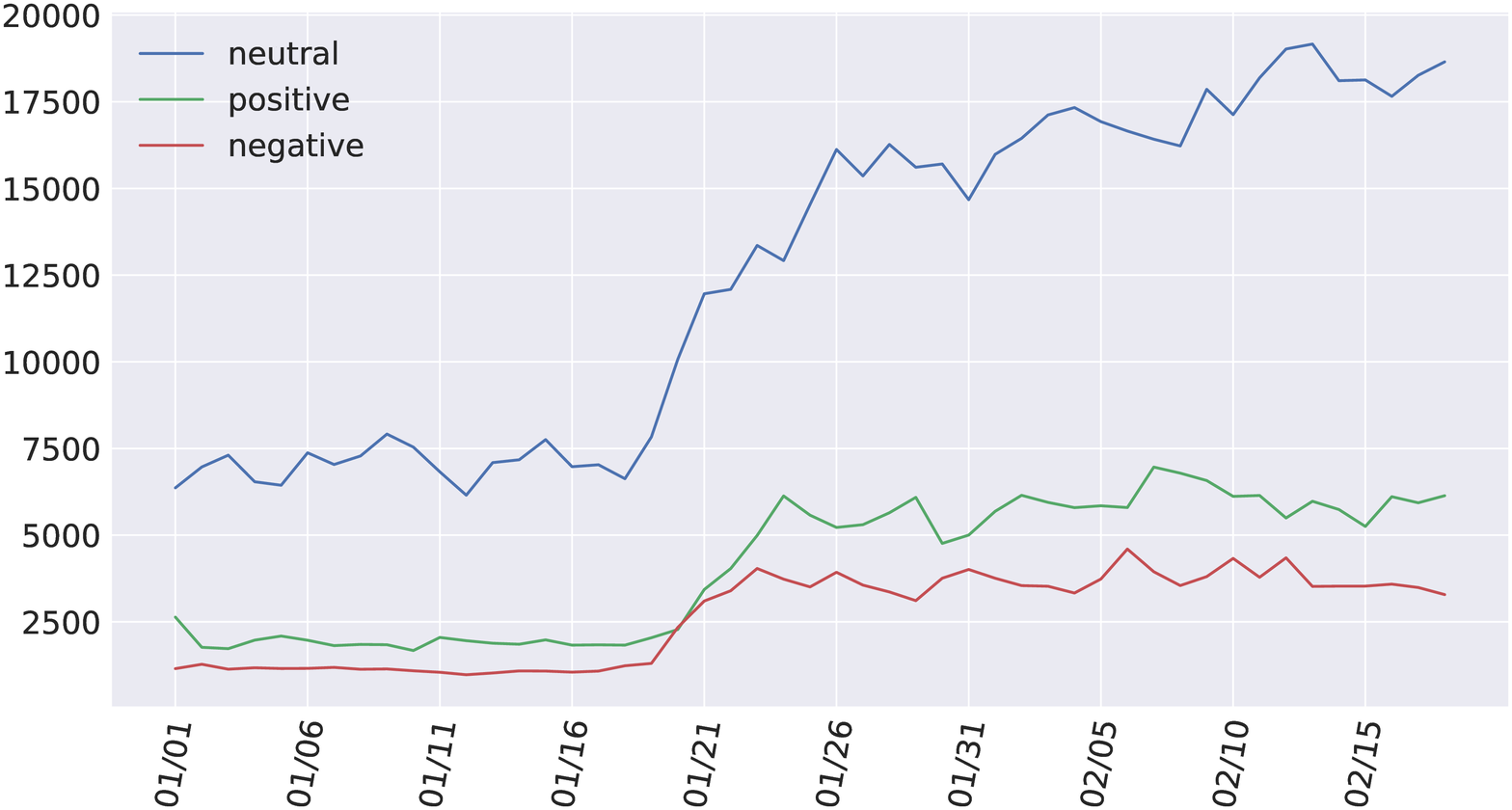}
  \vspace{-0.3cm}
  \caption{Number of microblogs of stage 1}
  \vspace{-0.3cm}
  \label{figure:stage1_line}
\end{figure}
\begin{figure}[t]
  \includegraphics[width=0.42\textwidth]{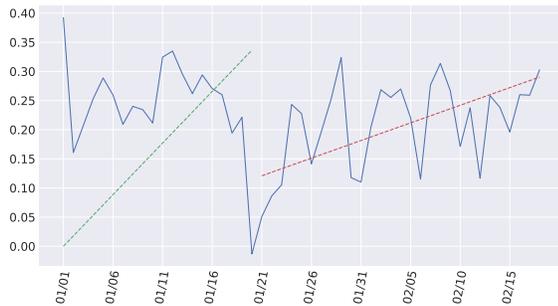}
  \vspace{-0.3cm}
  \caption{Sentiment Index of stage 1}
  \vspace{-0.3cm}
  \label{figure:stage1_si}
\end{figure}
Figure \ref{figure:stage1_sentiment} shows the sentiment proportion and Figure \ref{figure:stage1_line} shows the number of different sentiments from Jan. 1 to Feb. 18.
A direct observation is that most of the microblogs hold a neutral attitude towards the pandemic.
Considering the polarity of the opinions, there is a significant decline of the proportion of positive microblogs from Jan. 19 to Jan. 25.
Also, most of the microblogs were posted after Jan. 19.
Figure \ref{figure:stage1_si} shows the Sentiment Index from Jan. 1 to Feb. 18 and a significant decline could be observed near Jan. 20.

Based on the timeline, we can find two related key events:
(1) COVID-19 was announced to be Human-to-human transmissible on Jan. 20.
(2) A quarantine of the Greater Wuhan area beginning on Jan. 23 was announced on Jan. 22.
The influence of the these key events on public opinion is clear.
We regress the Sentiment Index against the number of days from Jan. 1 on the two parts divide by Jan. 21,  respectively, and report the regression coefficients (coef.) and t-statistics (t) as:
(1) Part-1. $coef.= 0.0177; t=5.169;  P>|t|:0.000$; and 
(2) Part-2. $coef.= 0.006; t=16.533;  P>|t|:0.000$.
Overall, the opinion was positive towards the pandemic and the sentiment was becoming positive after the decline.

\begin{figure}[t]
  \includegraphics[width=0.42\textwidth]{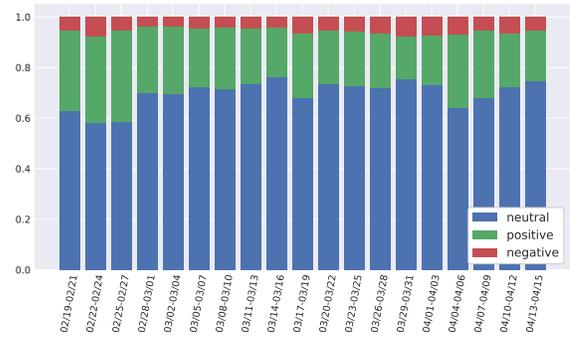}
  \vspace{-0.3cm}
  \caption{Sentiment proportion of stage 2}
  \vspace{-0.3cm}
  \label{figure:stage2_sentiment}
\end{figure}
\begin{figure}[t]
  \includegraphics[width=0.42\textwidth]{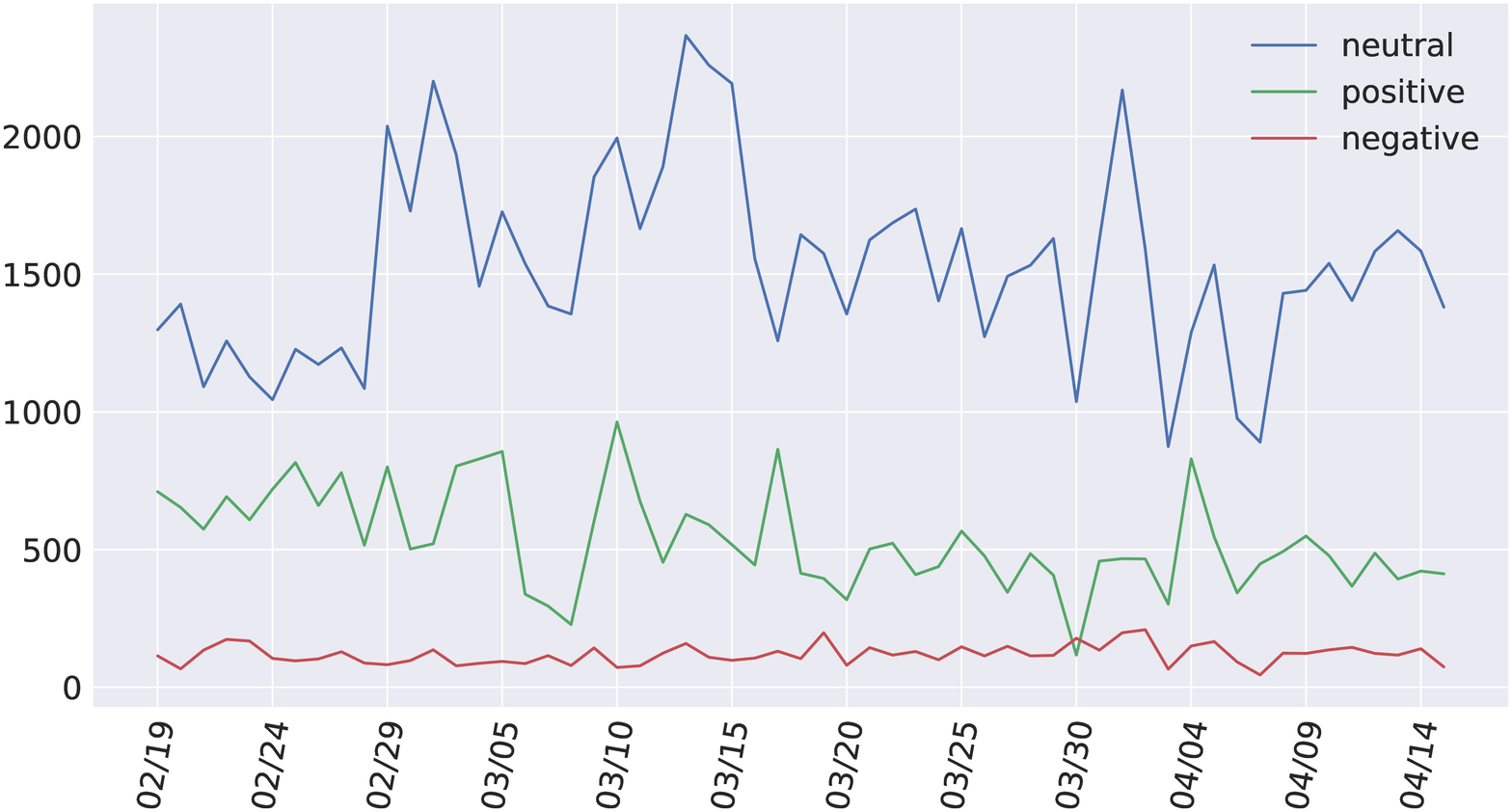}
  \vspace{-0.3cm}
  \caption{Number of microblogs of stage 2}
  \vspace{-0.3cm}
  \label{figure:stage2_line}
\end{figure}
\begin{figure}[t]
  \includegraphics[width=0.42\textwidth]{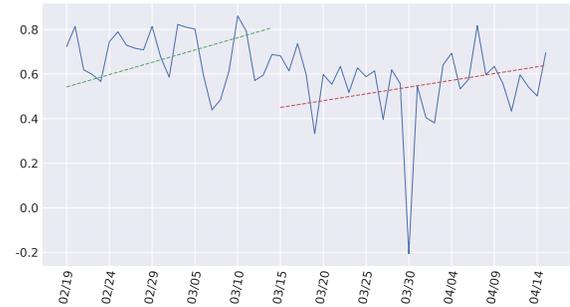}
  \vspace{-0.3cm}
  \caption{Sentiment Index of stage 2}
  \vspace{-0.3cm}
  \label{figure:stage2_si}
\end{figure}
Figure \ref{figure:stage2_sentiment} shows the sentiment proportion and Figure \ref{figure:stage2_line} shows the volume from Feb. 19 to Apr. 15.
A decrease of positive sentiment proportion can be observed from Feb. 28 to Mar. 1.
Based on the timeline, we can find the related events: First death was confirmed in U.S.
From figure \ref{figure:stage2_line} and \ref{figure:stage2_sentiment}, we can find that there is a decrease of the number of sentiment-positive microblogs near Mar. 15. 
The key event near Mar. 15 is that the confirmed cases in the U.S. increased from 1,000 to more than 10,000 during Mar. 10 to Mar. 19.
In addition, the U.S. President Donald Trump called novel coronavirus the `China virus' on Twitter on Mar. 16.
Based on this, Figure \ref{figure:stage2_si} shows the Sentiment Index from Feb. 19 to Apr. 15 and the two stages are divided by Mar. 15.
We regress the Sentiment Index against the number of days from Jan. 1 on the two parts respectively and report the regression coefficients (coef.) and t-statistics (t) as:
(1) Part-1. $coef.= 0.0111; t=22.4;  P>|t|:0.000$;
(2) Part-2. $coef.= 0.0061; t=16.923;  P>|t|:0.000$.
Basically, in this stage, the overall public sentiment was improving slowly and the second part is lower than the first part.
On the whole, positive microblogs are more than negative microblogs most of the time, while there is an obvious negative Sentiment Index near Mar. 30.
On that way two COVID-19 survivors beat the CT technician of a hospital, which ignited much discussion on Weibo.
\begin{figure}[h]
  \includegraphics[width=0.42\textwidth]{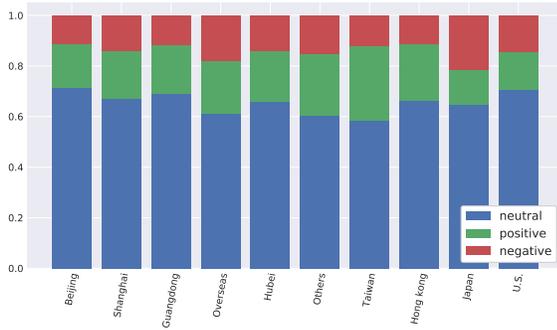}
  \vspace{-0.3cm}
  \caption{Sentiment in different regions}
  \vspace{-0.2cm}
  \label{figure:sentiment_province}
\end{figure}
\vspace{-0.3cm}
\vspace{-0.2cm}
\subsubsection{Public opinion of users from different regions}
We present region-related sentiment in this section.
First, we select several representative regions and show the number of microblogs with different types of sentiment in Figure \ref{figure:sentiment_province}. Clearly, Hong Kong and Taiwan hold more positive microblogs than negative microblogs.
The numbers of positive and negative microblogs are close from overseas and the U.S.
Japan posts more negative microblogs than positive microblogs.

We further present a detailed analysis on the relationship between sentiment and GDP per capita of a given province of China.
We rank the GDP per capita of Chinese provinces (except for Hong Kong, Macau, and Taiwan) and their positive/negative sentiment proportions.
To compare the two ranks, we use Normalized Spearman's footrule given by:
\begin{equation}
NFr(r_1,r_2) = 1 - \frac{Fr^{|S|}(r_1,r_2)} {max\quad Fr^{|S|}}
\end{equation}
£¬where $r_1,r_2$ are two permutations and $|S|$ is the number of overlapping items between them, when $|S|$ is odd $max\quad Fr^{|S|}=1/2(|S|+1)(|S|-1)$ and when $|S|$ is even $max\quad Fr^{|S|}=1/2|S|^2$.
$Fr^{|S|}(r_1,r_2)$  represents standard Spearman's footrule as:
\begin{equation}
Fr^{|S|}(r_1,r_2) = \sum_{i=1}^{|S|}|r_1(i) - r_2(i)|
\end{equation}
$NFr(r_1,r_2)$ ranges from 0 to 1 and a higher score indicates $r_1$ and $r_2$ are more similar and the comparison result of different lists is shown in Table  \ref{table:nfr}.
With the results of NFr, we can draw a preliminary conclusion that the higher GDP per capita a province has, the more negative microblogs and fewer positive microblogs it has.

\begin{table}[t]
\begin{tabular}{|c|c|}
\hline
ranks &  NFr  \\  \hline
positive (from high to low) \& GDP rank  &  0.13   \\ \hline
positive (from low to high) \& GDP rank  &   0.55 \\ \hline
negative (from high to low) \& GDP rank  &  0.53  \\ \hline
negative (from low to high) \& GDP rank  &  0.18  \\ \hline
\end{tabular}
\vspace{0.2cm}
\caption{NFr between different ranks of sentiment}
\label{table:nfr}
\vspace{-.4cm}
\end{table}

\begin{figure}[t]
  \includegraphics[width=0.42\textwidth]{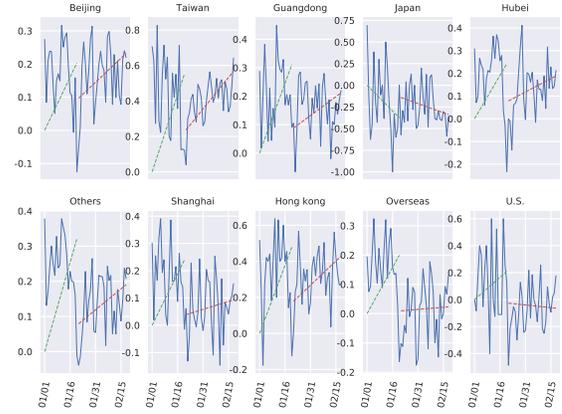}
  \vspace{-0.3cm}
  \caption{Sentiment Index by different regions of stage 1}
  \label{figure:stage1_si_province}
\end{figure}

Figure \ref{figure:stage1_si_province} shows the Sentiment Index in different regions.
The Sentiment Index is regressed against the number of days from Jan. 1 on the two parts divided by Jan. 21 respectively. The results of regression are shown in  Table \ref{table:reg_stage1}.
Most of the results pass the t-test expect for the U.S. and the part 2 of Japan.
There are several observations from Figure \ref{figure:stage1_si_province}:
(1) most of the regions held a positive attitude towards the pandemic before Jan. 21 and there was an clear decline on Jan. 21 like the overall sentiment in section \ref{Public opinion on different stages}.
Also, the gradient of the regression equation for most regions is higher in the first part than the second part;
(2) Hubei suffered a significant decline near Jan. 21 and the Sentiment Index was close to -0.2 here.
Overseas and the U.S. hold a similar pattern, especially the U.S., the lowest Sentiment Index of the U.S. is close to -0.4; and 
(3) Japan holds a declining regression equation in part 1, which differentiates it from other regions.

\begin{figure}[t]
  \includegraphics[width=0.42\textwidth]{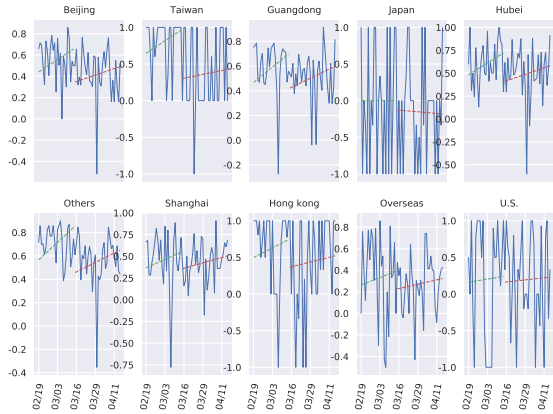}
  \caption{Sentiment Index by different regions of stage 2}
  \label{figure:stage2_si_province}
\end{figure}

Figure \ref{figure:stage2_si_province} shows the Sentiment Index in different regions on stage 2.
Sentiment Index is regressed against the number of days from Jan. 1 on the two parts divided by Mar. 15 respectively and the results of regression are shown in  Table \ref{table:reg_stage2}.
\begin{table}[t]
\small
\begin{tabular}{|c|c|c|c|c|c|c|}
\hline
\multirow{2}{*}{Region} & \multicolumn{3}{c|}{Part 1} & \multicolumn{3}{c|}{Part 2} \\\cline{2-7}
                  &  coef.     &    t   &   $P>|t|$   &     coef.  &    t   &  $P>|t|$    \\\hline
                  Beijing&  0.0107     &    3.338   &   0.003   &   0.0048    &  9.743     &  0.000    \\\hline
      Others        &      0.0169 &     4.702  &  0.000     &   0.0040    &    9.375   &     0.000          \\\hline
        Taiwan          &  0.029   &   3.777    &   0.001   &   0.0118    & 20.934      &  0.000    \\\hline
        Shanghai          &  0.0126     &   3.679    &   0.002   &    0.0019   &    3.408   &   0.002   \\\hline
         Guangdong         &   0.0164    &   5.120    &  0.000    &   0.0044    &   8.641    &   0.000  \\\hline
         Hong kong    &  0.0252    &   4.348    &    0.000   &   0.0089   &   11.637    &    0.000         \\\hline
         Japan &   -0.0197    &    -2.440   &   0.025   &   -0.007    & -5.744      & 0.000     \\\hline
         Overseas &    0.0105   &   3.399    &  0.003    &   0.0005    &  0.917     & 0.367     \\\hline
         Hubei &    0.0127   &   3.243    &  0.004    &  0.0039     &   6.869    &   0.000  \\\hline
         U. S. &    0.0110   &   1.937    & 0.068     &   -0.0013    &    -1.426   &  0.165    \\\hline
\end{tabular}
\caption{Regression analysis of stage 1}
\vspace{-0.3cm}
\label{table:reg_stage1}
\end{table}
\begin{table}[t]
\small
\begin{tabular}{|c|c|c|c|c|c|c|}
\hline
\multirow{2}{*}{Region} & \multicolumn{3}{c|}{Part 1} & \multicolumn{3}{c|}{Part 2} \\\cline{2-7}
                  &  coef.     &    t   &   $P>|t|$   &     coef.  &    t   &  $P>|t|$    \\\hline
                  Beijing&  0.0090     &    12.380   &   0.000   &   0.0047   &  9.563     &  0.000    \\\hline
      Others        &      0.0116 &    19.504  &  0.000     &   0.0063   &   13.806  &     0.000         \\\hline
        Taiwan          &  0.0132   &   11.406    &   0.000   &   0.0041    & 3.836      &  0.001    \\\hline
        Shanghai          &  0.0075     &   6.414    &   0.000   &    0.0048  &    10.972 &   0.000   \\\hline
         Guangdong         &   0.0095   &   12.134    & 0.000    &   0.0057   &   14.054    &  0.000 \\\hline
         Hong kong    &  0.0102    &   5.746    &    0.000   &   0.0050   &4.037   &    0.000       \\\hline
         Japan &   0.0000    &    -0.003   &   0.998  &   -0.0017   & -1.314     & 0.198    \\\hline
         Overseas &    0.0054   &   3.990   & 0.001   &   0.0031   & 6.242    & 0.000    \\\hline
         Hubei &    0.0099   &   12.805   & 0.000    &  0.0055     &   9.217   &   0.000  \\\hline
         U. S. &    0.0033  &   1.268    & 0.217     &   0.0022   &    1.811   &  0.080 \\\hline\end{tabular}
\caption{Regression analysis of stage 2}
\label{table:reg_stage2}
\end{table}
We can make several intuitive observations from the figure \ref{figure:stage2_si_province}.
(1) Microblogs from Japan and the U.S. are not enough to support a regression analysis.
There is no significant pattern that the sentiment of these two regions changed over time;
(2) An obvious decline can be observed near Mar. 30 in some regions like Beijing, and Shanghai. The hospital fighting event was mentioned in Section \ref{Public opinion on different stages}; and 
(3) There is a decline near Mar. 7 in several Chinese regions like Shanghai, Guangdong and outside China regions like Overseas and the U.S.
Two events can be found near Mar. 7:
Xinjia Express Hotel which served as a centralized medical observation point collapsed in Quanzhou, Fujian on Mar. 7.
COVID-19 infected Nicola Zingaretti, chairman of Partito Democratico.
\vspace{-0.2cm}
\subsubsection{Public opinion of users of different genders}
Considering users of different genders whether their microblogs are positive or negative, there are 25.2\% positive and 12.9\% negative in stage 1 and 33.0\% positive and 6.0\% negative in stage 2 for female users.
For male users, there are 16.5\% positive and 9.9\% negative in stage 1 and 16.6\% positive and 4.5\% negative in stage 2.
Most male and female users hold a neutral position and the proportions of positive and negative are close in both stages.
What is different is that a higher proportion of male users post neutral microblogs in stage 1.
The ratio of male to female microblogs is 81\%, that means more microblogs are posted by female.
An interesting finding is that in stage 2 the ratio of male to female microblogs is 1.06, which indicates with the development of pandemic, the proportion of microblogs by male users is increasing.
\vspace{-0.2cm}
\subsubsection{Public opinion of users with different age}
\begin{figure}[t]
  \includegraphics[width=0.4\textwidth]{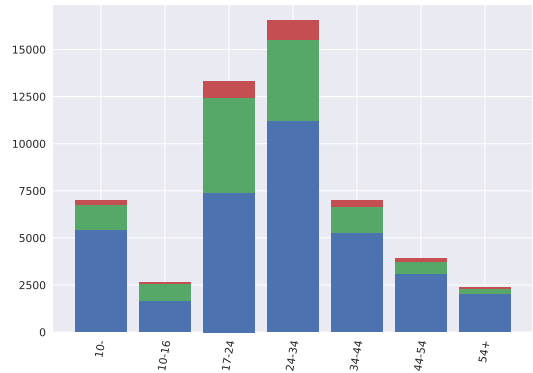}
  \vspace{-0.3cm}
  \caption{Sentiment of different users}
  \vspace{-0.3cm}
  \label{figure:age}
\end{figure}

Only user profiles from stage 2 provide information about their birthdays, allowing us to analyze the users in stage 2 by considering their age. The result is shown in Figure \ref{figure:age}.
Most microblogs were posted by users from 17 to 34, while most of the positive and negative microblogs were posted by them at the same time.
Users from 17 to 34 prefer to express their positive and negative opinions.
\vspace{-0.2cm}
\subsubsection{Public opinion of users with different educational background}
\begin{figure}[t]
\begin{minipage}[b]{0.49\linewidth}
\centering
\includegraphics[width=0.99\textwidth]{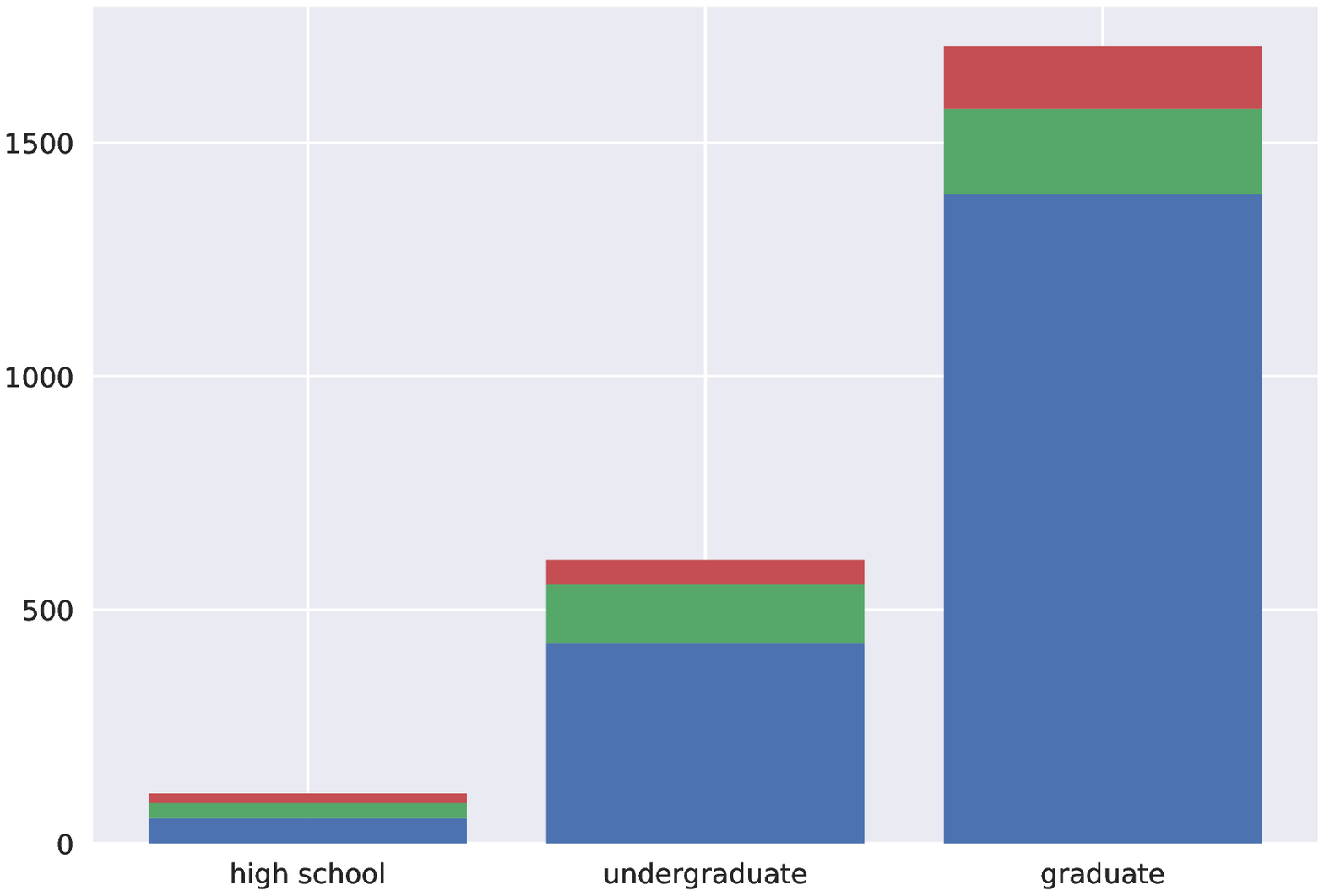}
\vspace{-0.3cm}
\centerline{(a) Stage 1}
\end{minipage}
\begin{minipage}[b]{0.49\linewidth}
\centering
\includegraphics[width=0.99\textwidth]{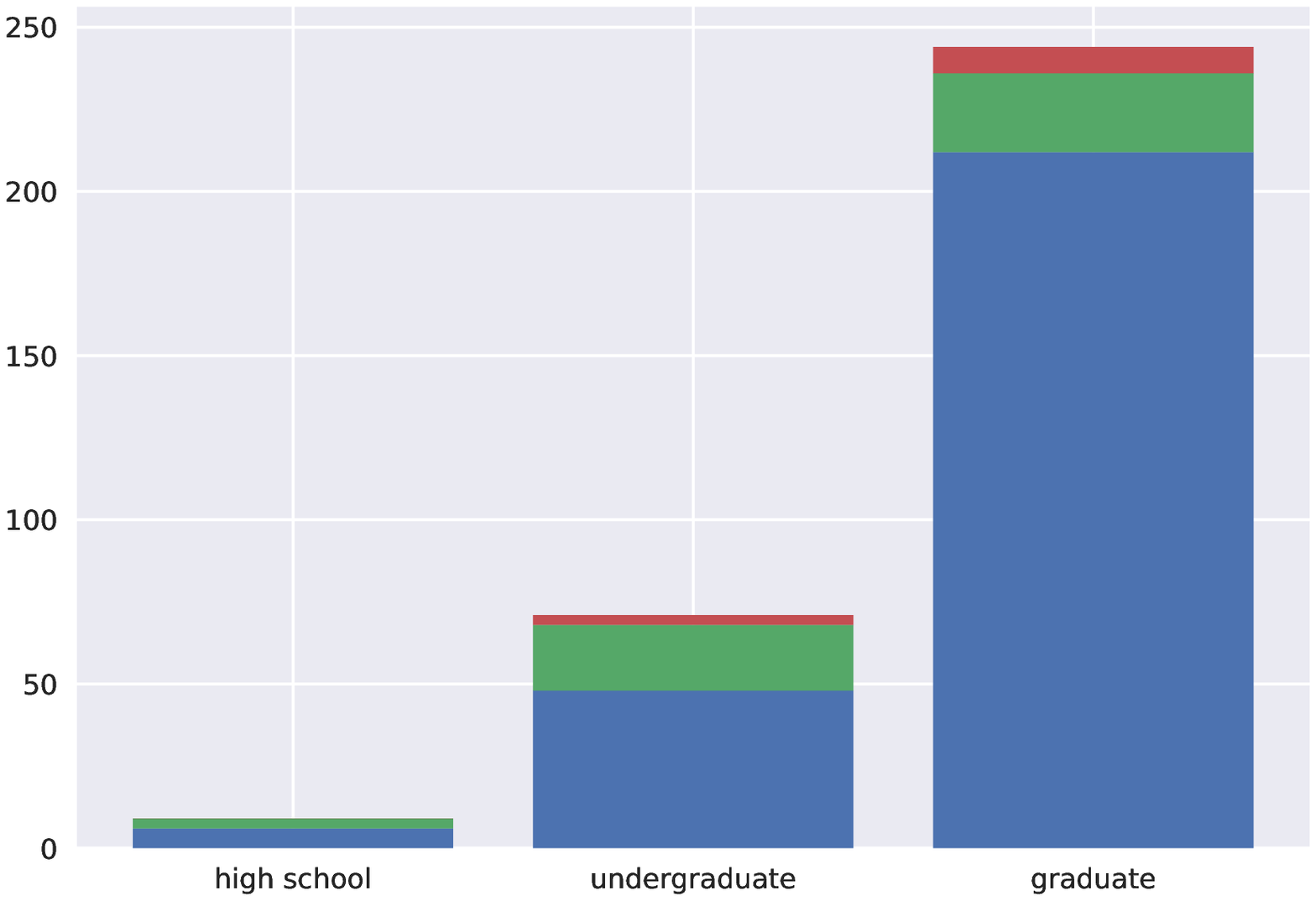}
\vspace{-0.3cm}
\centerline{(b) Stage 2}
\end{minipage}
\caption{Sentiment of different users on different stages}
\vspace{-0.3cm}
 \label{figure:edu}
 \end{figure}

Few users provide their educational background. We filter the educational background of a specific user by searching key words like 'high school student' in the brief introduction of their profiles.
With the results shown in \ref{figure:edu}, we can find that microblogs with higher educational background are more likely to be negative.
 
\section{Specific Topics}
\begin{table}[h]
\small
\begin{tabular}{|c|c|c|c|c|c|c|}
\hline
\multirow{2}{*}{Topic} & \multicolumn{3}{c|}{Part 1} & \multicolumn{3}{c|}{Part 2} \\\cline{2-7}
                  &  coef.     &    t   &   $P>|t|$   &     coef.  &    t   &  $P>|t|$    \\\hline
                  China&  0.0526   &     6.161    &   0.000       &   0.0191  &    25.709   &   0.000    \\\hline
                  U.S.  &      -0.0135  &   -1.186   &      0.250      &  -0.0213  &   -16.993  &     0.000         \\\hline
\end{tabular}
\vspace{0.2cm}
\caption{Regression analysis on the public opinion of China and the U.S. during stage 1}
\vspace{-0.1cm}
\label{table:stage1_si_gov}
\end{table}
\subsection{China and the U.S. related microblogs}
China and the U.S. are two regions of high interest. 
We first make an analysis on the volume of microblogs related to the two topics on different stages.
It is shown that 11.5\% microblogs discussing China and 0.9\% microblogs discussing U.S. on stage 1 and on stage 2 there are 18.7\% microblogs for China and 8.4\% for U.S.
We can see a significant increase in proportion of microblogs discussing U.S. and China comparing different stages of the COVID-19 pandemic.

\begin{figure}[t]
  \includegraphics[width=0.42\textwidth]{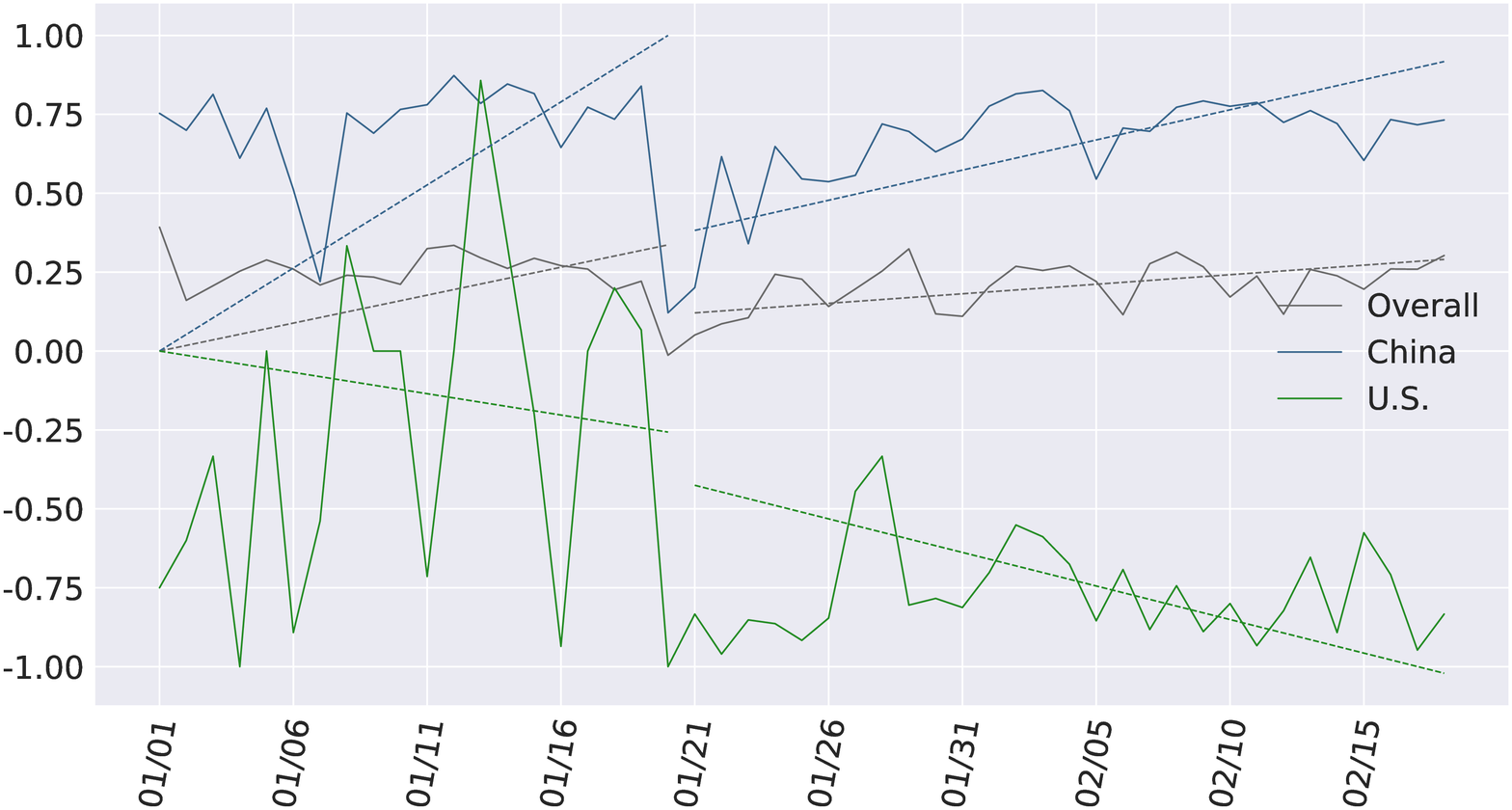}
  \caption{Sentiment Index China and the U.S.}
  \label{figure:stage1_si_gov}
\end{figure}

The Sentiment Indices of the microblogs from Jan. 1 to Feb. 18 discussing China and the U.S. are shown in Figure \ref{figure:stage1_si_gov} and the regression statistics are shown in Table \ref{table:stage1_si_gov}.
We can make several intuitive observations:
(1) In general, the public attitude towards China was more positive than towards the U.S.
(2) During part 1, the public opinion on U.S. was fluctuating and slumped after Jan. 21.

\begin{figure}[t]
  \includegraphics[width=0.42\textwidth]{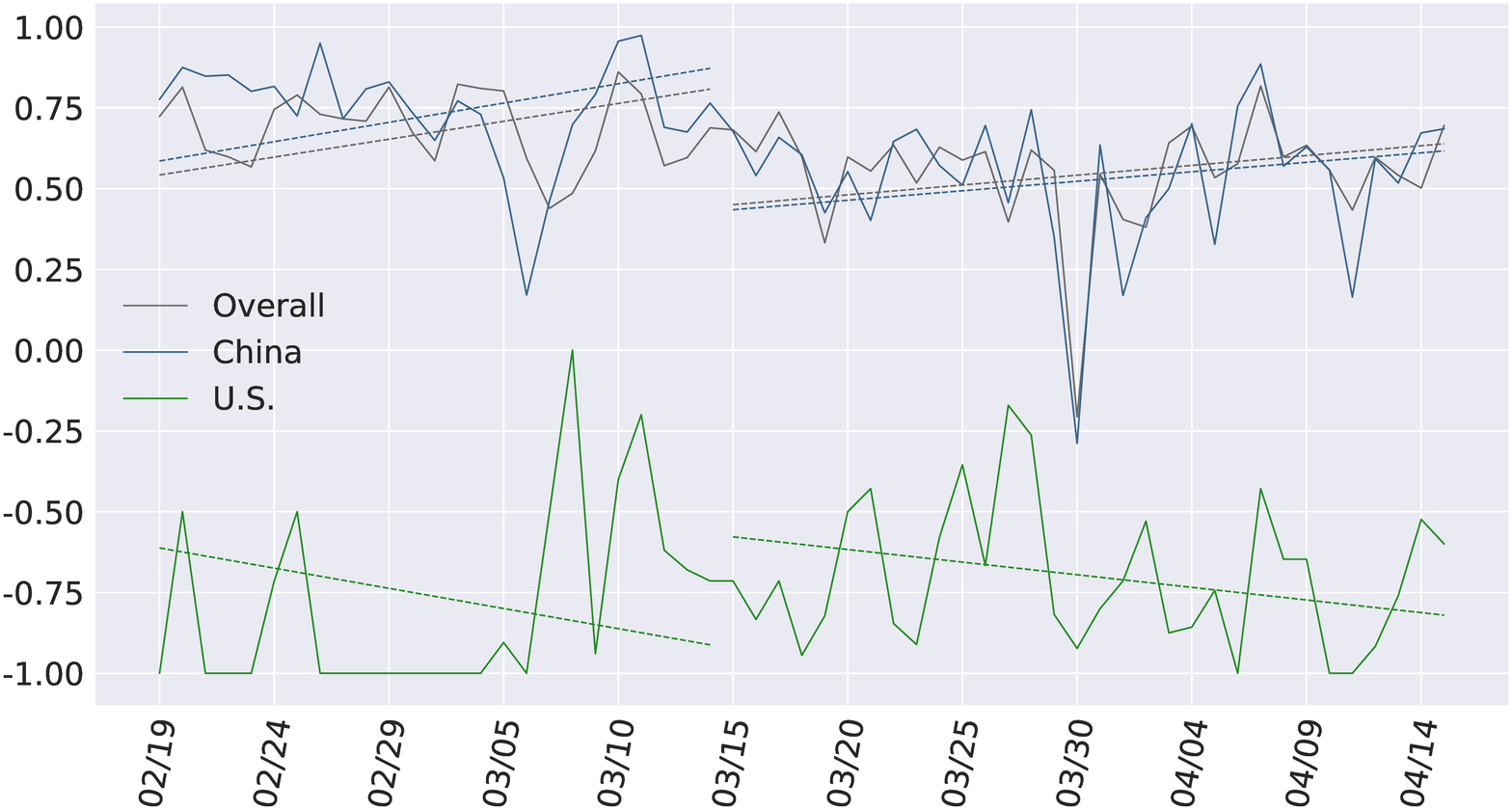}
  \caption{Sentiment Index on China and the U.S.}
  \label{figure:stage2_si_gov}
\end{figure}

The Sentiment Index and corresponding regression statistics on Microblogs from Feb. 19 to Apr. 15 discussing the Chinese government and U.S. government are shown in Figure \ref{figure:stage2_si_gov} and Table \ref{table:stage2_si_gov}.
It is shown that the public opinion on the Chinese government is similar with overall opinion on the pandemic, while the public attitude towards U.S. government is below them.
\begin{table}[h]
\small
\begin{tabular}{|c|c|c|c|c|c|c|}
\hline
\multirow{2}{*}{Topic} & \multicolumn{3}{c|}{Part 1} & \multicolumn{3}{c|}{Part 2} \\\cline{2-7}
                  &  coef.     &    t   &   $P>|t|$   &     coef.  &    t   &  $P>|t|$    \\\hline
                  China.&   0.0119   &     17.557  &   0.000       &   0.0059    &     13.203    &   0.000    \\\hline
                  U.S..    &       -0.0125   &    -11.249      &      0.250      &   -0.0078    &    -17.829    &     0.000         \\\hline
\end{tabular}
\vspace{0.2cm}
\caption{Regression analysis on the public opinion of China and the U.S. during stage 2}
\vspace{0.2cm}
\label{table:stage2_si_gov}
\end{table}

\begin{table}[h]
\small
\begin{tabular}{|c|c|c|}
\hline
\multirow{2}{*}{Sentiment Index} & \multicolumn{2}{c|}{correlation coefficient} \\ \cline{2-3}
                            &       stage 1     &      stage 2     \\ \hline
          Overall \&  China   &      0.62    &      0.79     \\ \hline
          Overall \&  U.S.    &       0.26    &        -0.02   \\ \hline
          China   \& U.S.     &        0.39   &       0.06    \\ \hline
\end{tabular}
\caption{Correlation Coefficient on the Sentiment Index of different topics}
\vspace{-0.3cm}
\label{table:corr_gov}
\end{table}
\vspace{-0.3cm}
We further validate the relationship between the public opinion on China, the U.S. and overall public opinion with Pearson Correlation Coefficients and the results are shown in Table \ref{table:corr_gov}.
The highest correlations are achieved by overall and China in stage 2 and we can find satisfied results on overall and China in both stages.
In addition, it is noticeable that the coefficient between the microblogs of China and the U.S. in stage 1 is 0.39.

\begin{figure}[t]
  \includegraphics[width=0.42\textwidth]{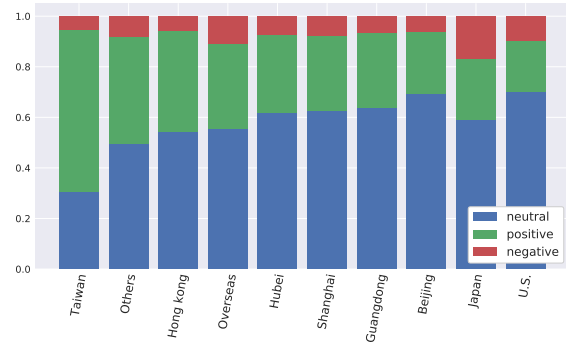}
  \vspace{-0.3cm}
  \caption{On the Chinese government in different regions}
  \vspace{-0.3cm}
  \label{figure:sentiment_province_cn_gov_all}
\end{figure}

\begin{figure}[t]
  \includegraphics[width=0.42\textwidth]{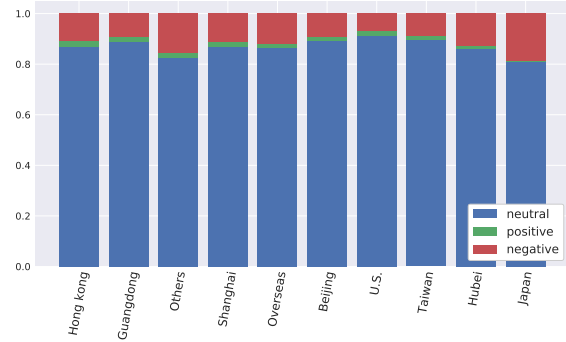}
  \vspace{-0.3cm}
  \caption{On the U.S. government in different regions}
  \vspace{-0.3cm}
  \label{figure:sentiment_province_us_gov_all}
\end{figure}

We also provide an analysis of the opinion on China and the U.S. by considering the  regions of users.
Figures \ref{figure:sentiment_province_cn_gov_all} and \ref{figure:sentiment_province_us_gov_all} show the results of sentiment proportions in different regions.
Considering the microblogs about China, only Japan holds a similar number of positive and negative microblogs.
When it comes to the microblogs about the U.S. in Figure \ref{figure:sentiment_province_us_gov_all}, there are more negative microblogs than positive microblogs in most regions.

In addition, we provide a further analysis on Chinese government-related and U.S. government-related microblogs.
Since the volume of government-related microblogs is not enough to make an analysis based on time, we provide a direct analysis on the volume.
Based on the statistics, the Sentiment Index in all stages for the topic `China' is $0.69$ and for the topic `U.S.' is $-0.72$, and the Sentiment Index on microblogs directly mentioning `Chinese government' is $0.09$ and that for `U.S. government' is $-0.96$.
It is shown that most microblogs show a negative attitude towards the U.S. and U.S. government, which means the public opinions on them are consistent.
In contrast, there is a significantly higher proportion of negative microblogs of the  Chinese government than China.
 \vspace{-0.1cm}
\subsection{Term Usage}
\begin{figure}[t!]
  \includegraphics[width=0.42\textwidth]{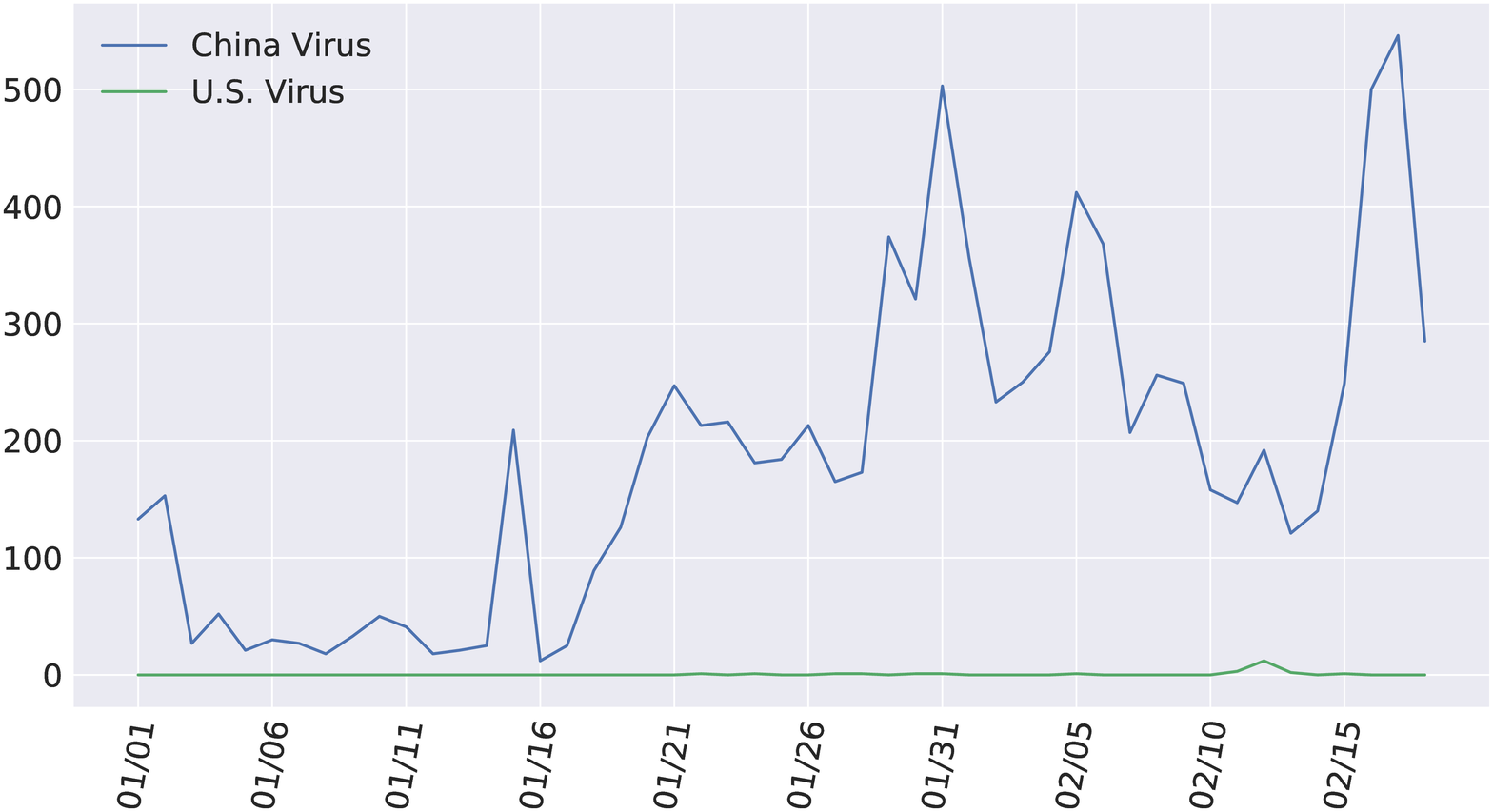}
  \vspace{-0.3cm}
  \caption{Term usage during stage 1}
  \vspace{-0.3cm}
  \label{figure:stage1_line_virus.eps}
\end{figure}

\begin{figure}[t!]
  \includegraphics[width=0.42\textwidth]{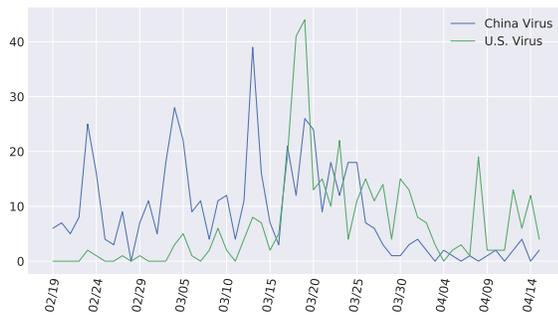}
  \vspace{-0.3cm}
  \caption{Term usage during stage 2}
  \vspace{-0.3cm}
  \label{figure:stage2_line_virus.eps}
\end{figure}

There are different types of terms referring to COVID-19 by users.
For example, controversial expressions which connect region and virus such as `China virus' and `U.S. virus' are used during the pandemic.
We show the usage of different terms during different stages of the pandemic in Figure \ref{figure:stage1_line_virus.eps} and Figure \ref{figure:stage2_line_virus.eps}
It is clear that `U.S. virus' were  used more in stage 2.
Considering the Sentiment Index on the different topics of the two stages, `China virus' is -0.50 and -0.68.
When it comes to `U.S. virus', the Sentiment Index in stage 2 is -0.89.

That means the Chinese public expressed negative sentiment when using these terms in general.
Also, some peaks were influenced by the China-US relationship.
For example, on Mar. 19 the CNN reporters noticed that the `corona virus' in the U.S. President's speech was manually changed to the word `Chinese virus', an immediate reaction by using `U.S. virus' can be observed near Mar. 19.

\subsection{Daily life during the COVID-19 pandemic}
\begin{figure*}[htbp]
\begin{minipage}[t]{0.3\linewidth}
\centering
\includegraphics[width=5cm]{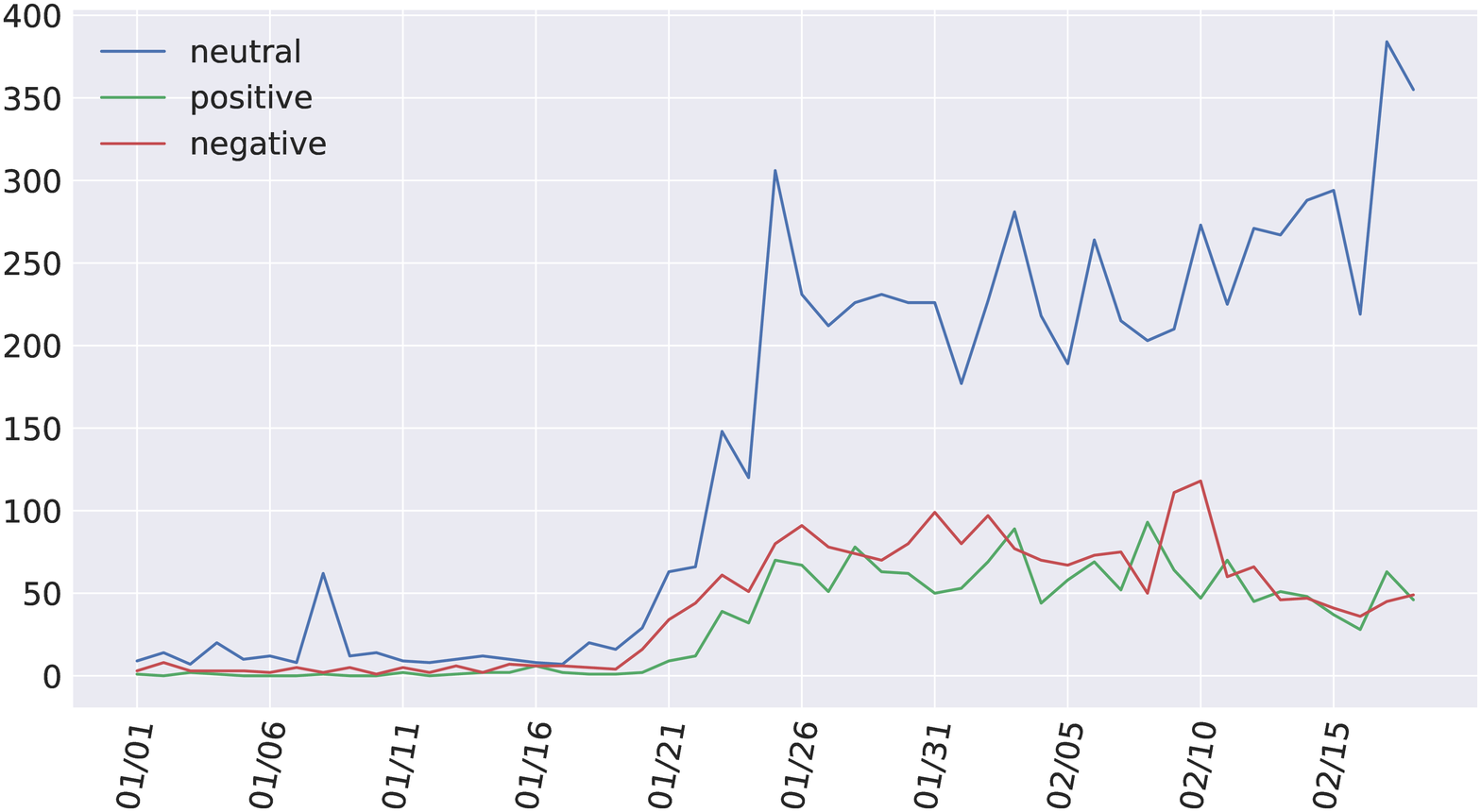}
  \vspace{-0.1cm}
  \caption{Number of microblogs on staying at home during stage 1}
    \vspace{-0.2cm}
  \label{figure:stage1_line_home}
\end{minipage}
\begin{minipage}[t]{0.3\linewidth}
\centering
 \includegraphics[width=5cm]{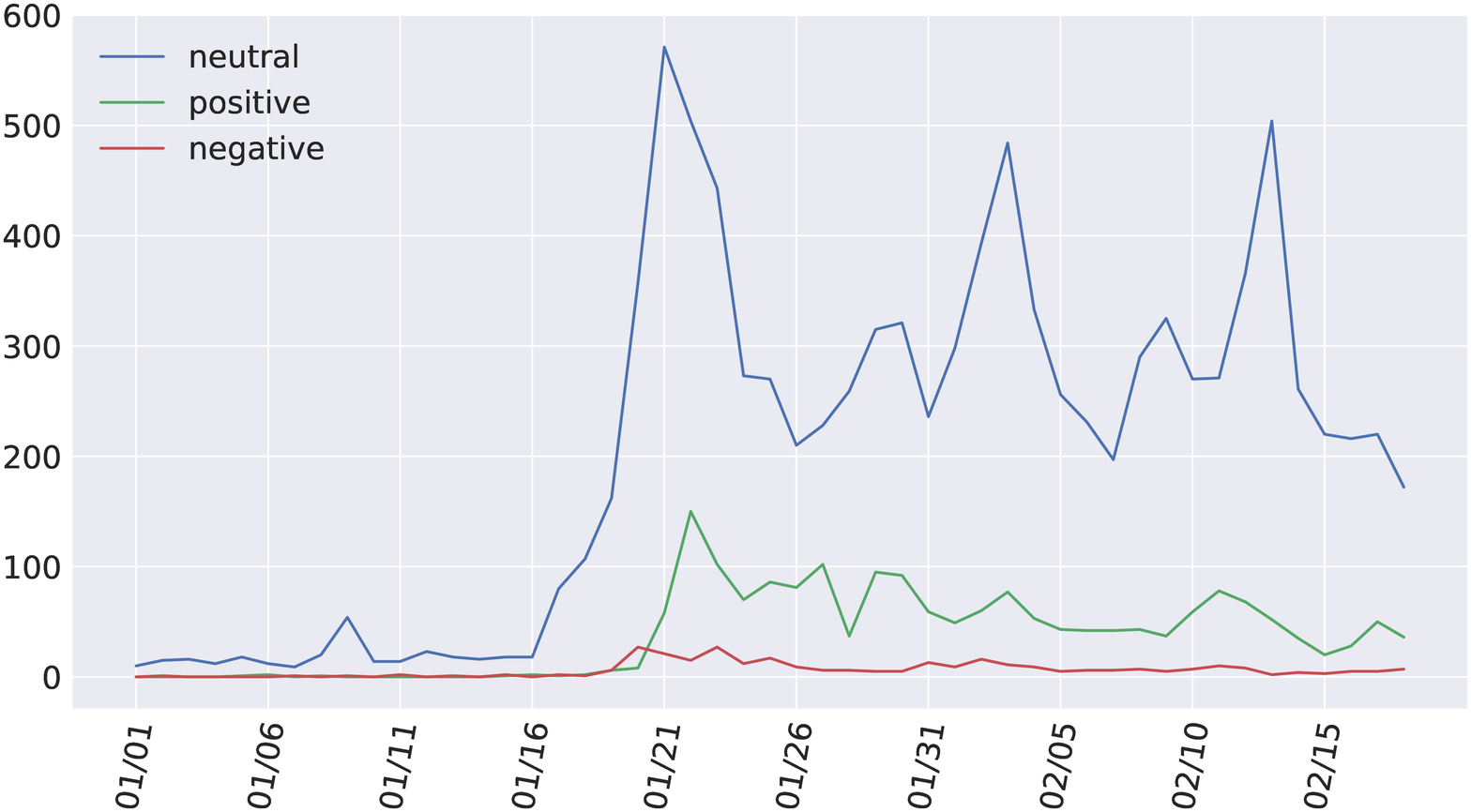}
   \vspace{-0.1cm}
  \caption{Number of microblogs on washing hands during stage 1}
    \vspace{-0.2cm}
  \label{figure:stage1_line_wash}
\end{minipage}
\begin{minipage}[t]{0.3\linewidth}
\centering
\includegraphics[width=5cm]{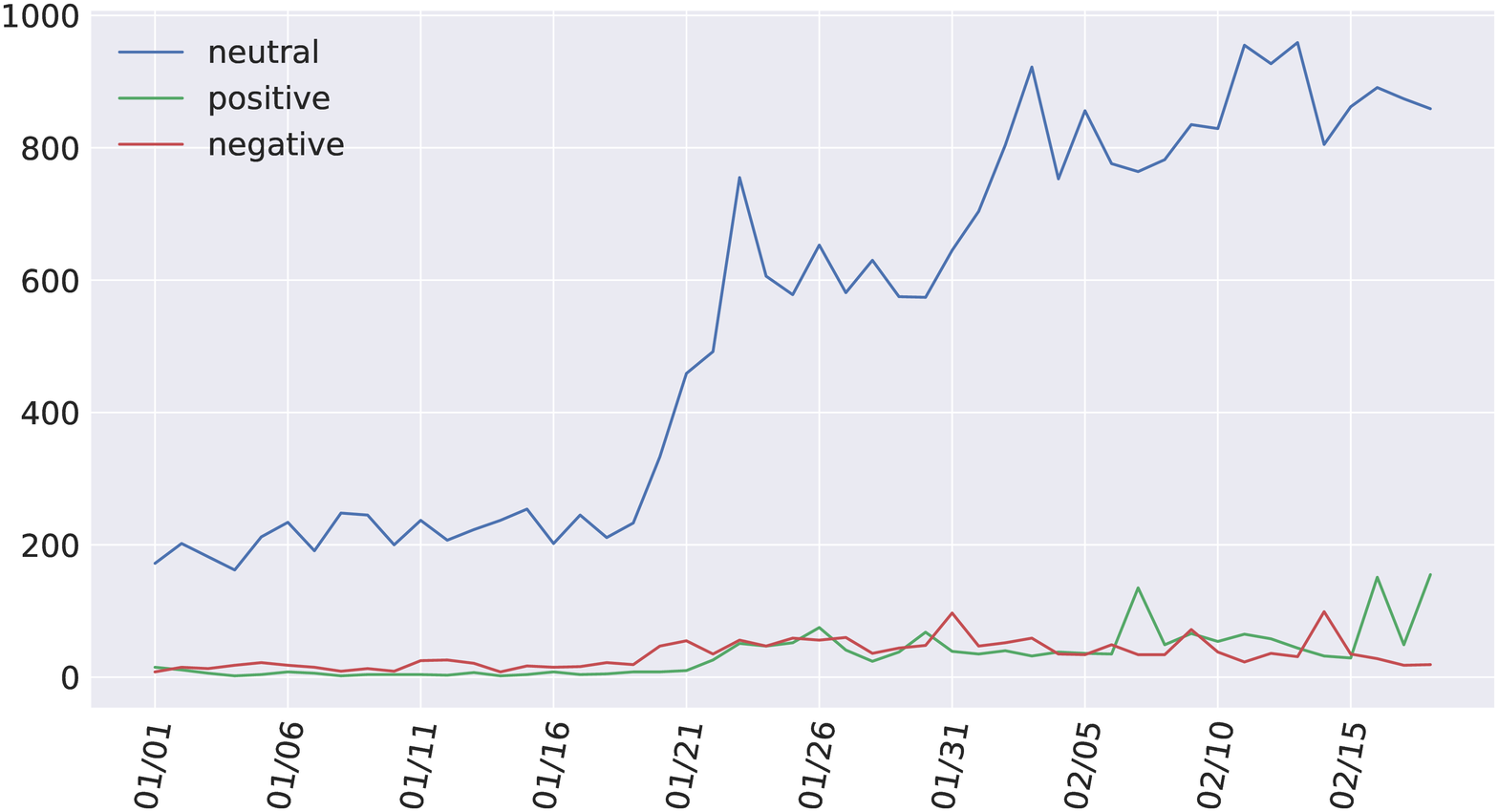}
  \vspace{-0.1cm}
  \caption{Number of microblogs on disinfection during stage 1}
    \vspace{-0.2cm}
  \label{figure:stage1_line_dis}
\end{minipage}%
\end{figure*}

\begin{figure*}[htbp]
\begin{minipage}[t]{0.3\linewidth}
\centering
  \includegraphics[width=5cm]{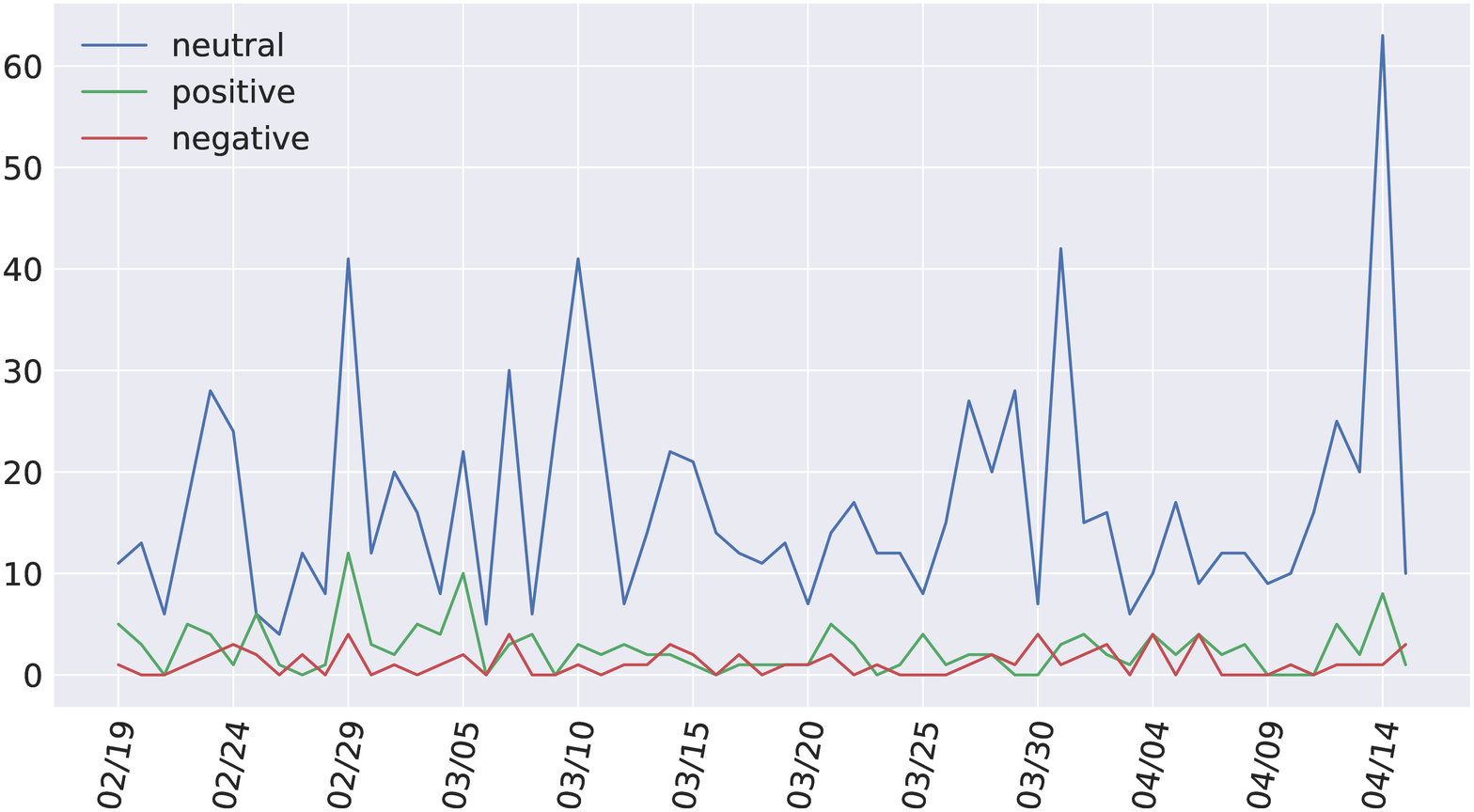}
  \vspace{-0.1cm}
  \caption{Number of microblogs on staying at home during stage 2}
  \vspace{-0.2cm}
  \label{figure:stage2_line_home}
\end{minipage}
\begin{minipage}[t]{0.3\linewidth}
\centering
   \includegraphics[width=5cm]{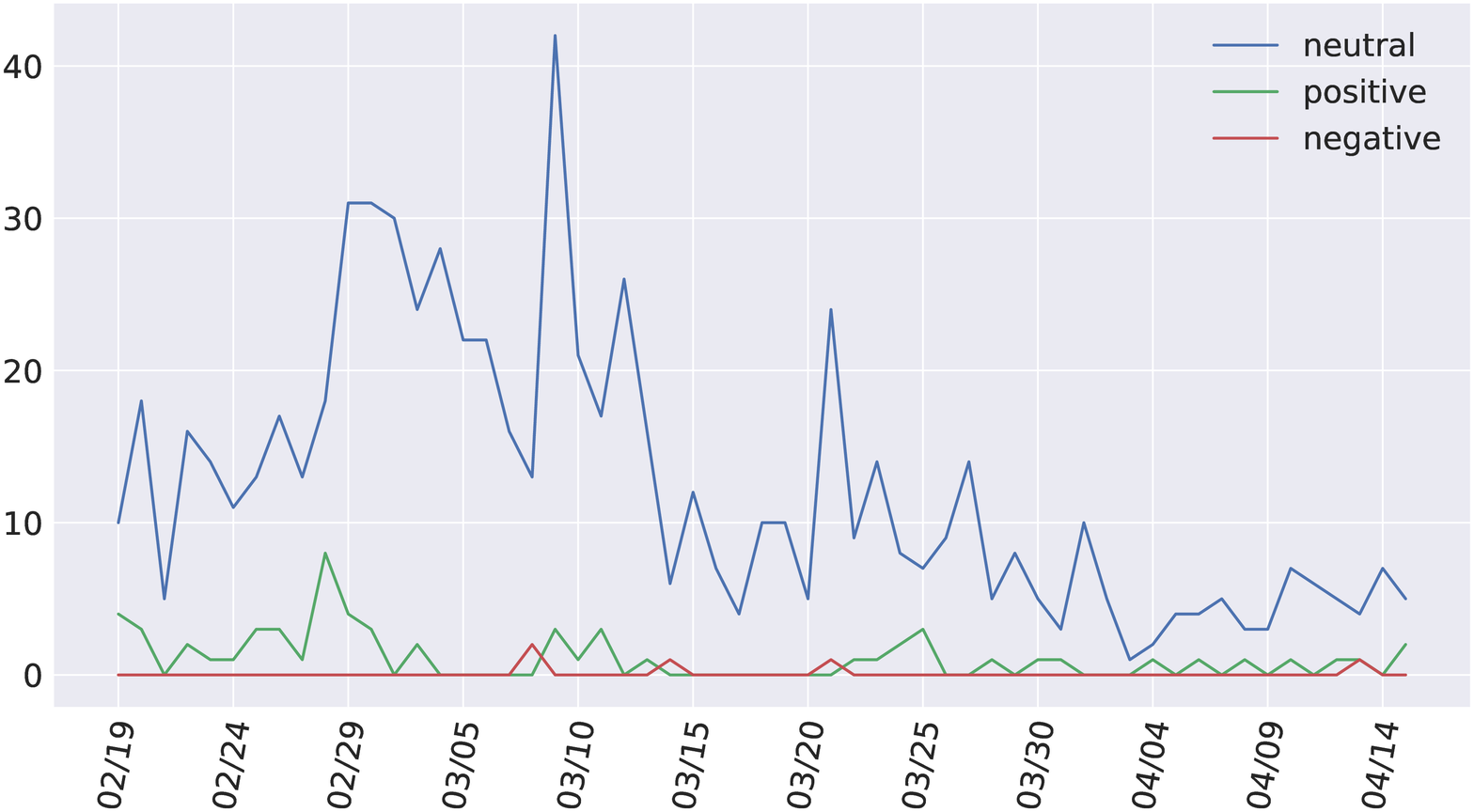}
   \vspace{-0.1cm}
  \caption{Number of microblogs on  washing hands during stage 2}
  \vspace{-0.2cm}
  \label{figure:stage2_line_wash}
\end{minipage}
\begin{minipage}[t]{0.3\linewidth}
\centering
\includegraphics[width=5cm]{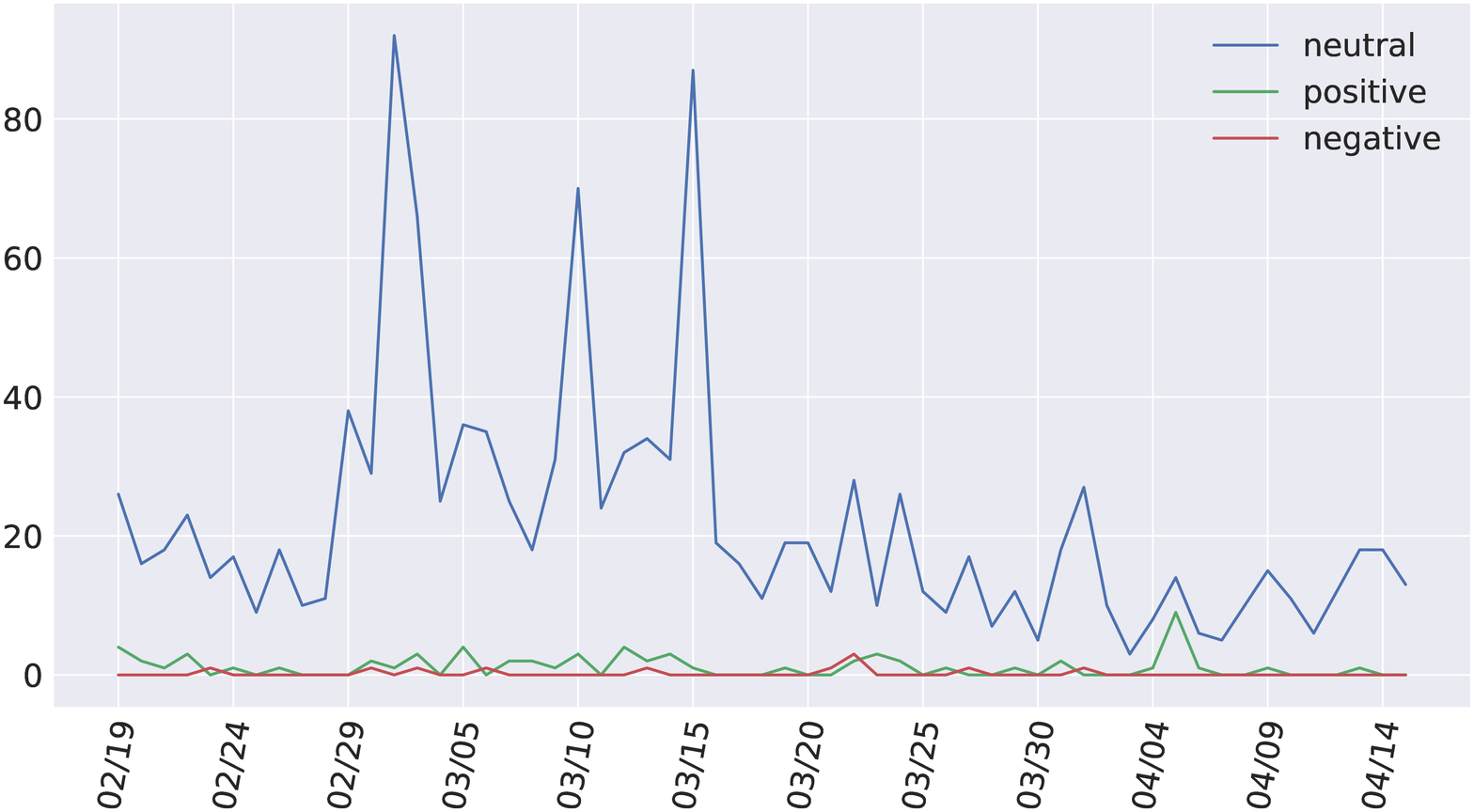}
\vspace{-0.1cm}
  \caption{Number of microblogs on disinfection during stage 2}
  \vspace{-0.2cm}
  \label{figure:stage2_line_dis}
\end{minipage}%
\end{figure*}

\begin{figure*}[t]
\begin{minipage}[t]{0.3\linewidth}
\centering
\includegraphics[width=5cm]{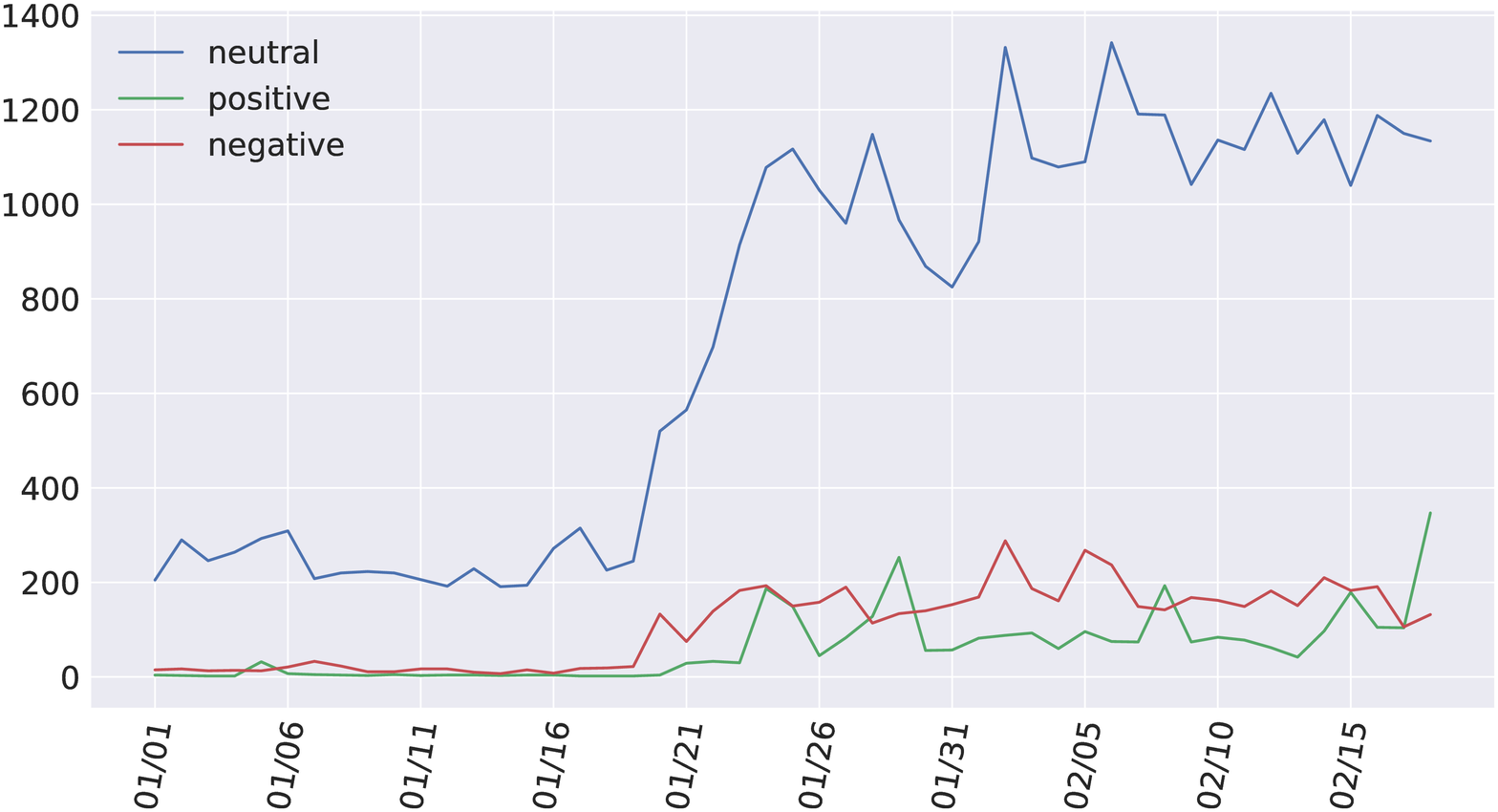}
\vspace{-0.1cm}
  \caption{Number of microblogs on quarantine during stage 1}
  \vspace{-0.2cm}
  \label{figure:stage1_line_iso}
\end{minipage}
\begin{minipage}[t]{0.3\linewidth}
\centering
 \includegraphics[width=5cm]{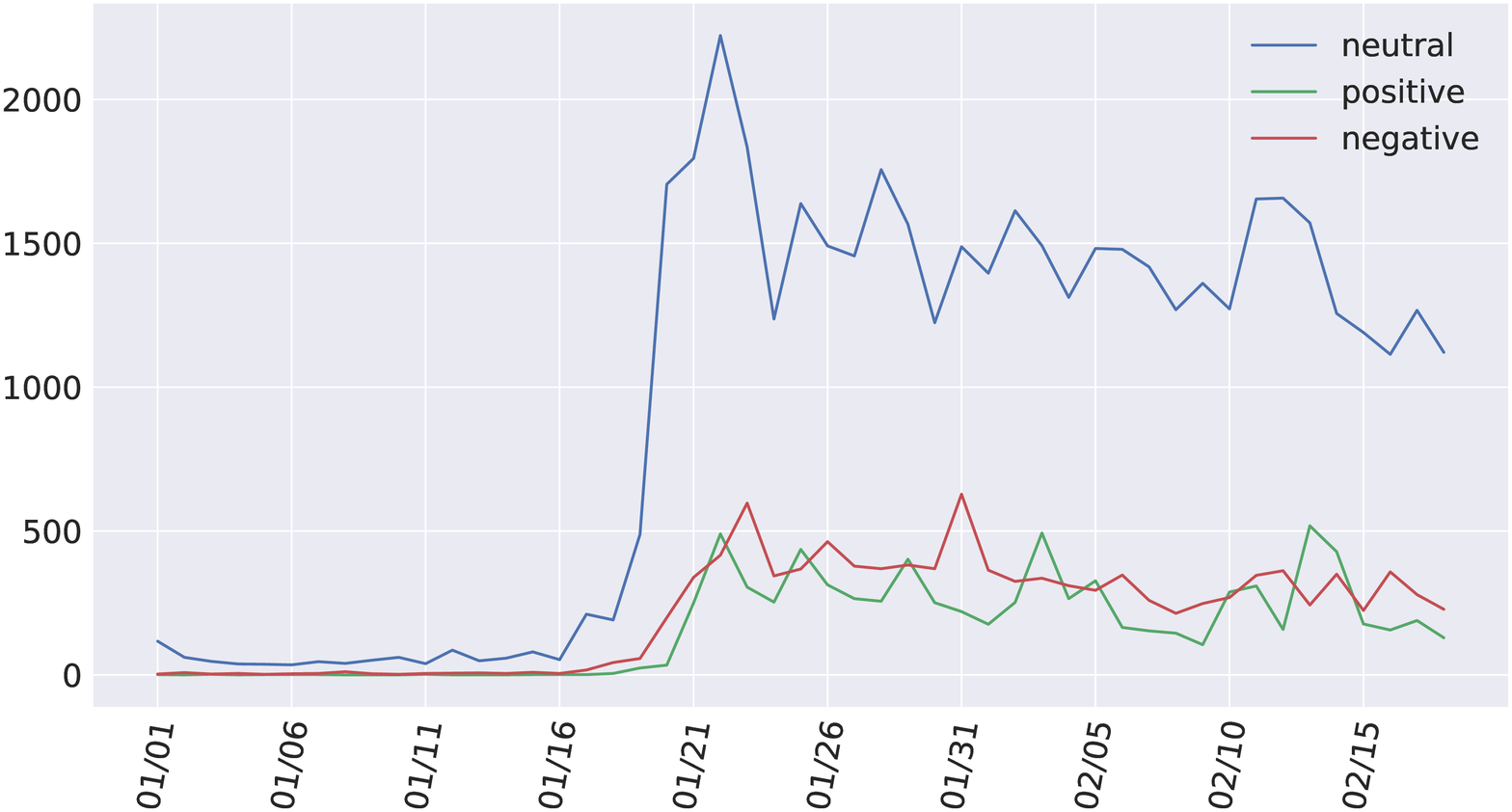}
 \vspace{-0.1cm}
  \caption{Number of microblogs on mask during stage 1}
  \vspace{-0.2cm}
  \label{figure:stage1_line_mask}
\end{minipage}
\begin{minipage}[t]{0.3\linewidth}
\centering
\includegraphics[width=5cm]{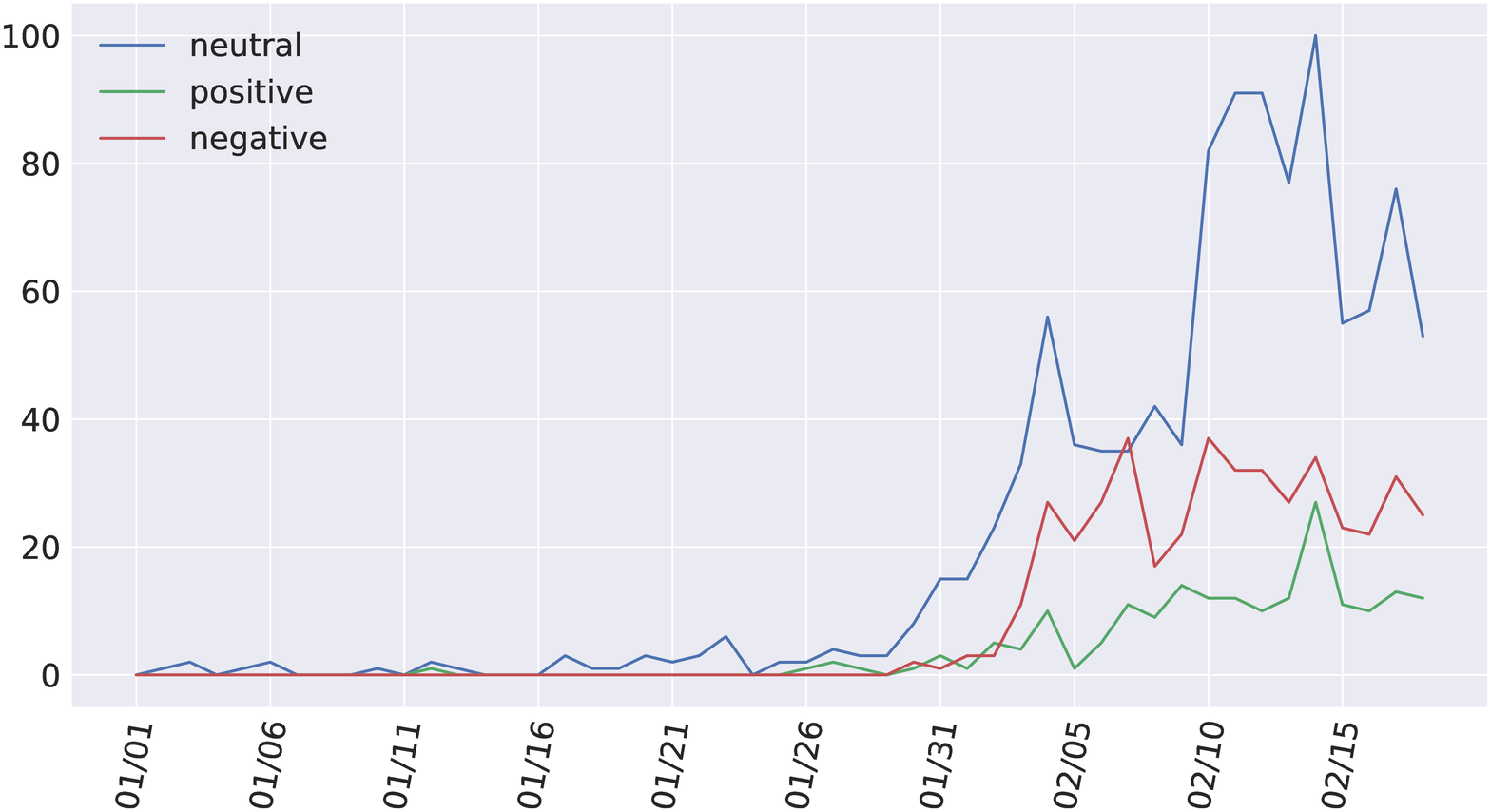}
\vspace{-0.1cm}
  \caption{Number of microblogs on online learning during stage 1}
  \vspace{-0.2cm}
  \label{figure:stage1_line_online}
\end{minipage}%
\end{figure*}

\begin{figure*}[t]
\begin{minipage}[t]{0.3\linewidth}
\centering
  \includegraphics[width=5cm]{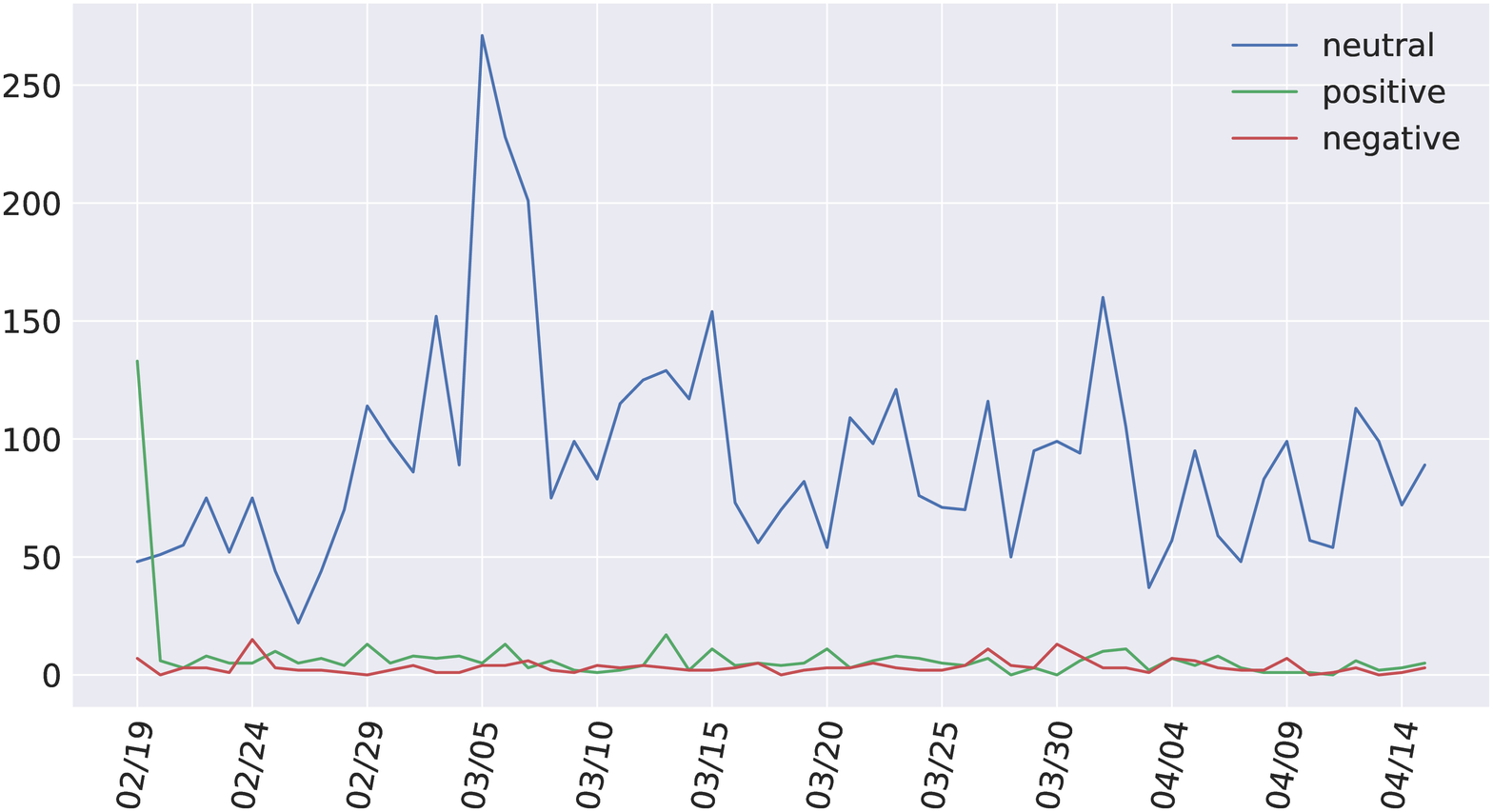}
  \vspace{-0.1cm}
  \caption{Number of microblogs on quarantine during stage 2}
  \vspace{-0.2cm}
  \label{figure:stage2_line_iso}
\end{minipage}
\begin{minipage}[t]{0.3\linewidth}
\centering
   \includegraphics[width=5cm]{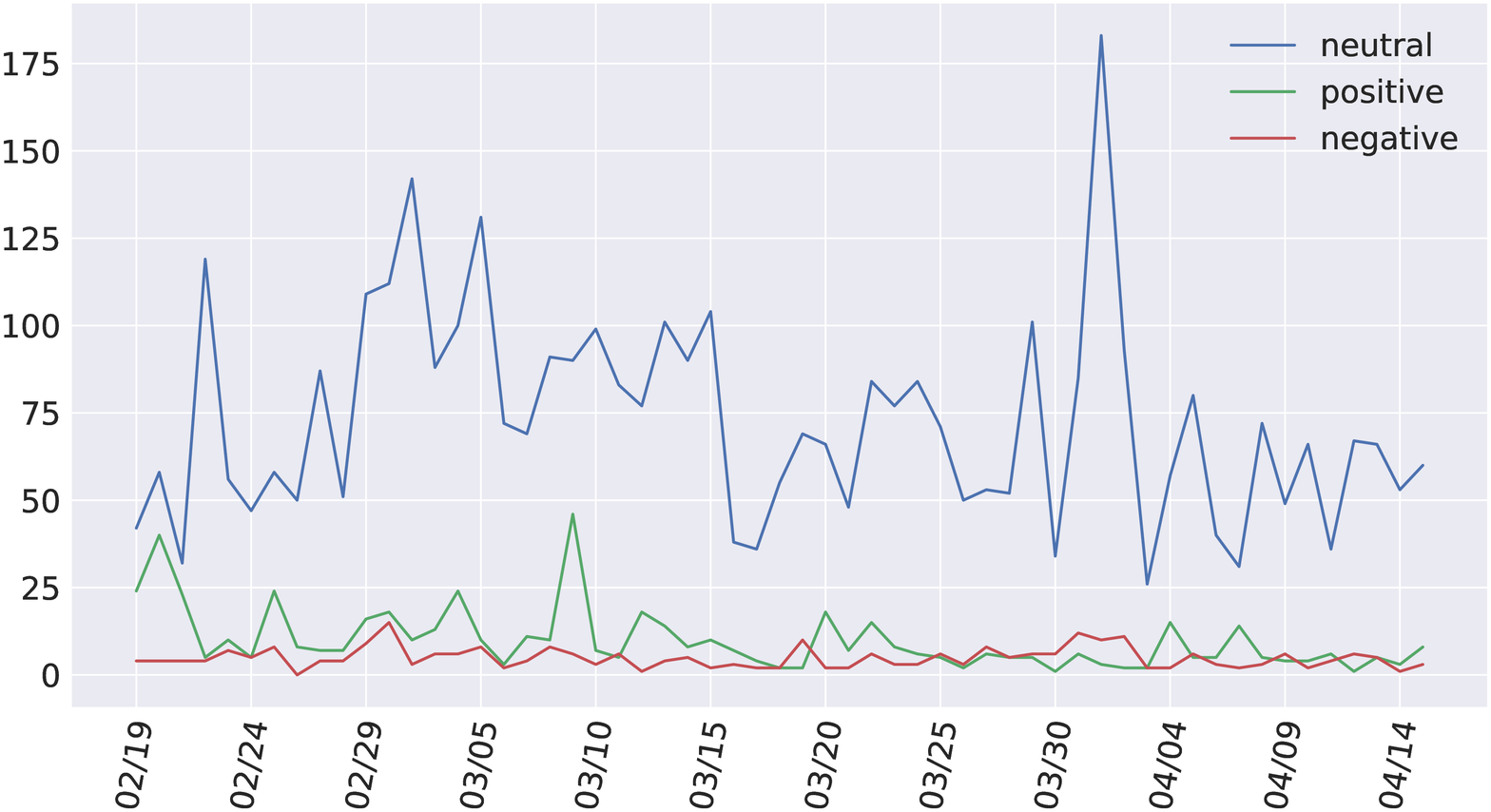}
   \vspace{-0.1cm}
  \caption{Number of microblogs on mask during stage 2}
  \vspace{-0.2cm}
  \label{figure:stage2_line_mask}
\end{minipage}
\begin{minipage}[t]{0.3\linewidth}
\centering
 \includegraphics[width=5cm]{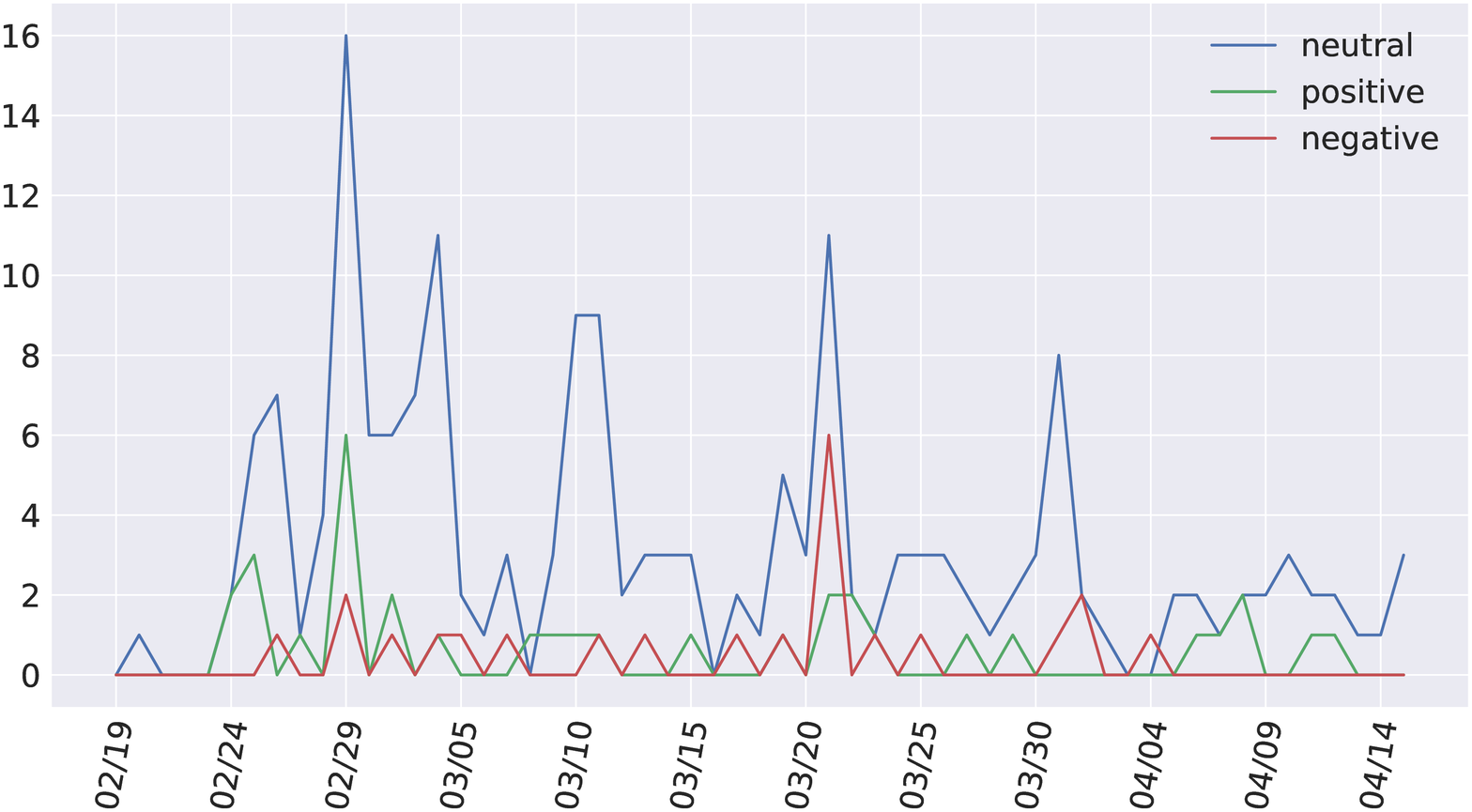}
 \vspace{-0.1cm}
  \caption{Number of microblogs on online learning during stage 2}
  \vspace{-0.2cm}
  \label{figure:stage2_line_online}
\end{minipage}%
\end{figure*}

Next, we discuss several common topics of daily life during the COVID-19 pandemic: staying at home, washing hands, disinfection, quarantine, mask and online learning.
There are some similarities as well as differences among them and we will discuss them as shown from Figure \ref{figure:stage1_line_home} to Figure \ref{figure:stage2_line_online}.

\vspace{-0.2cm}
\subsubsection{General Observations}
On most topics about daily life during the COVID-19 pandemic, discussion increased near Jan. 21. 
As mentioned before, COVID-19 was officially announced to be human-to-human transmissible on Jan. 20. and there would be a quarantine of the Great Wuhan region beginning on Jan. 23.
Therefore, a few days before and after Jan. 21 are the key dates when discussions on different aspects of daily life influenced by COVID-19 picked up, except for online learning.

Some peaks that appeared in stage 2 are also shared among different topics from Feb. 29 to Mar. 15.
There are several key events during this period: 
On Feb. 29, the U.S. reported the first death case of COVID-19; on Mar. 10, the confirmed cases in the U.S. increased to 1,000; and On Mar. 13 Trump issued the social distancing policy.

Some activities within the Weibo platform also influenced the discussions by users.
For example, many people shared a microblog such as `Don't party, go out less, wash your hands and wear masks! I am using \#Weibo avatar pendant\#, to fight the pandemic together, let's start from wearing a mask'.
Slogans like this were shared among Weibo users which increased the number of microblogs discussing daily life and caused the similar patterns on the volumes of microblogs discussing these topics.

Considering the sentiment polarity, most of the microblogs are neutral, and the numbers of  positive and negative microblogs are similar from the general opinion,  except for washing hands in Figure \ref{figure:stage1_line_wash} and online learning in Figure \ref{figure:stage1_line_online}.

For each of the  topics, we provide potential influential events and hot topics on Weibo to match the unique peak(s) of the topic.
\vspace{-0.2cm}
\subsubsection{Staying at home}
There is an obvious peak on Jan. 25 as shown in Figure \ref{figure:stage1_line_home}. On Jan. 25, one hashtag discussing `Cooking failures when staying at home' was widely shared on Weibo.
\vspace{-0.2cm}
\subsubsection{Quarantine}
On Mar. 3, 11 new imported COVID-19 cases were reported in Gansu Province.
\vspace{-0.2cm}
\subsubsection{Mask}
Hubei plans to request emergency support of masks and other medical supplies near Jan. 22.
Also, with the increasing demand of masks, people start to discuss how to buy masks, different types of masks and the price of masks which brings microblogs with negative information.
\vspace{-0.2cm}
\subsubsection{Online learning}
Feb. 03 is the first workday after Spring Festival, student start to discuss about online learning.
There are some hot hashtags like `\#Do not start online teaching before officially announced school opens\#' near Feb. 04.
Multiple provinces confirmed to delay school opening near Feb. 14.
\vspace{-0.2cm}
\subsubsection{Washing hands}
Peter Navarro warned again about COVID-19 in memorandum on Feb.3.
Most of microblogs mention washing hands encourage public to keep the good habit which brings positive emotions.
\vspace{-0.2cm}
\subsubsection{Disinfection}
Different to other topics, disinfection is discussed widely since Jan. 1.
Quarantine and disinfection are the only two topics which had been concerned by public from Jan. 1.
Most of the peaks of disinfection is similar to other topics.
\vspace{-0.2cm}
\subsection{Conclusion}
The COVID-19 pandemic has had an enormous impact on the world and we track the public opinion on Weibo during different stages of the pandemic.
Through the analysis of collected data, we find several factors that may influence the discussions on social media and public opinion:
(1) Different stages of the COVID-19 pandemic. It is clear that in different stages of the pandemic, the public opinion varied.
For example, when COVID-19 was officially announced human-to-human transmissible  the discussions on COVID-19 increased significantly.
(2) Policies. Major policies during the pandemic may ignite conversation, such as the Wuhan lock-down.
(3) China-US relationship. Use of `Chinese virus' by the U.S. President caused heated discussions.
(4) Infected celebrities. The news that the Chairman of Partito Democratico was infected gave rise to many microblogs about COVID-19 in Italy.
(5) User-generated topics about daily life during the pandemic. For example, a hashtag about cooking at home during the pandemic was widely used.
With this work, we provide a multi-faceted data analysis on the public opinion during different stages of COVID-19 pandemic on different topics. We hope more detailed analyses such as this can help understand the public reactions and prepare the public and governments for a prolonged COVID-19 pandemic or future pandemics.

\bibliographystyle{ACM-Reference-Format}
\bibliography{sample-sigconf}


\begin{thebibliography}{21}


\ifx \showCODEN    \undefined \def \showCODEN     #1{\unskip}     \fi
\ifx \showDOI      \undefined \def \showDOI       #1{#1}\fi
\ifx \showISBNx    \undefined \def \showISBNx     #1{\unskip}     \fi
\ifx \showISBNxiii \undefined \def \showISBNxiii  #1{\unskip}     \fi
\ifx \showISSN     \undefined \def \showISSN      #1{\unskip}     \fi
\ifx \showLCCN     \undefined \def \showLCCN      #1{\unskip}     \fi
\ifx \shownote     \undefined \def \shownote      #1{#1}          \fi
\ifx \showarticletitle \undefined \def \showarticletitle #1{#1}   \fi
\ifx \showURL      \undefined \def \showURL       {\relax}        \fi
\providecommand\bibfield[2]{#2}
\providecommand\bibinfo[2]{#2}
\providecommand\natexlab[1]{#1}
\providecommand\showeprint[2][]{arXiv:#2}

\bibitem[\protect\citeauthoryear{Agichtein, Castillo, Donato, Gionis, and
  Mishne}{Agichtein et~al\mbox{.}}{2008}]%
        {agichtein2008finding}
\bibfield{author}{\bibinfo{person}{Eugene Agichtein}, \bibinfo{person}{Carlos
  Castillo}, \bibinfo{person}{Debora Donato}, \bibinfo{person}{Aristides
  Gionis}, {and} \bibinfo{person}{Gilad Mishne}.}
  \bibinfo{year}{2008}\natexlab{}.
\newblock \showarticletitle{Finding high-quality content in social media}. In
  \bibinfo{booktitle}{\emph{Proceedings of the 2008 international conference on
  web search and data mining}}. \bibinfo{pages}{183--194}.
\newblock


\bibitem[\protect\citeauthoryear{Badawy, Ferrara, and Lerman}{Badawy
  et~al\mbox{.}}{2018a}]%
        {Badawy2018AnalyzingTD}
\bibfield{author}{\bibinfo{person}{Adam Badawy}, \bibinfo{person}{Emilio
  Ferrara}, {and} \bibinfo{person}{Kristina Lerman}.}
  \bibinfo{year}{2018}\natexlab{a}.
\newblock \showarticletitle{Analyzing the Digital Traces of Political
  Manipulation: The 2016 Russian Interference Twitter Campaign}.
\newblock \bibinfo{journal}{\emph{International Conference on Advances in
  Social Networks Analysis and Mining}} (\bibinfo{year}{2018}).
\newblock


\bibitem[\protect\citeauthoryear{Badawy, Ferrara, and Lerman}{Badawy
  et~al\mbox{.}}{2018b}]%
        {b15}
\bibfield{author}{\bibinfo{person}{Adam Badawy}, \bibinfo{person}{Emilio
  Ferrara}, {and} \bibinfo{person}{Kristina Lerman}.}
  \bibinfo{year}{2018}\natexlab{b}.
\newblock \showarticletitle{Analyzing the Digital Traces of Political
  Manipulation: The 2016 Russian Interference Twitter campaigns}. In
  \bibinfo{booktitle}{\emph{International Conference on Advances in Social
  Networks Analysis and Mining}}.
\newblock


\bibitem[\protect\citeauthoryear{Bergsma and Durme}{Bergsma and Durme}{2013}]%
        {Bergsma2013UsingCC}
\bibfield{author}{\bibinfo{person}{Shane Bergsma} {and}
  \bibinfo{person}{Benjamin~Van Durme}.} \bibinfo{year}{2013}\natexlab{}.
\newblock \showarticletitle{Using Conceptual Class Attributes to Characterize
  Social Media Users}. In \bibinfo{booktitle}{\emph{ACL}}.
\newblock


\bibitem[\protect\citeauthoryear{Cinelli, Quattrociocchi, Galeazzi, Valensise,
  Brugnoli, Schmidt, Zola, Zollo, and Scala}{Cinelli et~al\mbox{.}}{2020}]%
        {Cinelli2020TheCS}
\bibfield{author}{\bibinfo{person}{Matteo Cinelli}, \bibinfo{person}{Walter
  Quattrociocchi}, \bibinfo{person}{Alessandro Galeazzi},
  \bibinfo{person}{Carlo~Michele Valensise}, \bibinfo{person}{Emanuele
  Brugnoli}, \bibinfo{person}{Ana~Luc{\'i}a Schmidt}, \bibinfo{person}{Paola
  Zola}, \bibinfo{person}{Fabiana Zollo}, {and} \bibinfo{person}{Antonio
  Scala}.} \bibinfo{year}{2020}\natexlab{}.
\newblock \showarticletitle{The COVID-19 Social Media Infodemic}.
\newblock \bibinfo{journal}{\emph{ArXiv}} (\bibinfo{year}{2020}).
\newblock


\bibitem[\protect\citeauthoryear{Giahanou and Crestani}{Giahanou and
  Crestani}{2016}]%
        {Giahanou2016LikeIO}
\bibfield{author}{\bibinfo{person}{Anastasia Giahanou} {and}
  \bibinfo{person}{Fabio Crestani}.} \bibinfo{year}{2016}\natexlab{}.
\newblock \showarticletitle{Like It or Not: A Survey of Twitter Sentiment
  Analysis Methods}.
\newblock \bibinfo{journal}{\emph{ACM Comput. Surv.}}  \bibinfo{volume}{49}
  (\bibinfo{year}{2016}), \bibinfo{pages}{28:1--28:41}.
\newblock


\bibitem[\protect\citeauthoryear{Hutto and Gilbert}{Hutto and Gilbert}{2014}]%
        {hutto2014vader}
\bibfield{author}{\bibinfo{person}{Clayton~J Hutto} {and} \bibinfo{person}{Eric
  Gilbert}.} \bibinfo{year}{2014}\natexlab{}.
\newblock \showarticletitle{Vader: A parsimonious rule-based model for
  sentiment analysis of social media text}. In \bibinfo{booktitle}{\emph{Eighth
  international AAAI conference on weblogs and social media}}.
\newblock


\bibitem[\protect\citeauthoryear{Jin, Gallagher, Han, and Luo}{Jin
  et~al\mbox{.}}{2010}]%
        {j2}
\bibfield{author}{\bibinfo{person}{Xin Jin}, \bibinfo{person}{Andrew
  Gallagher}, \bibinfo{person}{Jiawei Han}, {and} \bibinfo{person}{Jiebo Luo}.}
  \bibinfo{year}{2010}\natexlab{}.
\newblock \showarticletitle{Wisdom of Social Multimedia: Using Flickr for
  Prediction and Forecast}. In \bibinfo{booktitle}{\emph{ACM Multimedia
  Conference}}.
\newblock


\bibitem[\protect\citeauthoryear{Lu, Hu, Wang, Kumar, Liu, and Maciejewski}{Lu
  et~al\mbox{.}}{2015}]%
        {Lu2015VisualizingSM}
\bibfield{author}{\bibinfo{person}{Yafeng Lu}, \bibinfo{person}{Xia Hu},
  \bibinfo{person}{Feng Wang}, \bibinfo{person}{Shamanth Kumar},
  \bibinfo{person}{Huan Liu}, {and} \bibinfo{person}{Ross Maciejewski}.}
  \bibinfo{year}{2015}\natexlab{}.
\newblock \showarticletitle{Visualizing Social Media Sentiment in Disaster
  Scenarios}. In \bibinfo{booktitle}{\emph{WWW}}.
\newblock


\bibitem[\protect\citeauthoryear{Maas, Daly, Pham, Ng, and Potts}{Maas
  et~al\mbox{.}}{2011}]%
        {Maas2011LearningWV}
\bibfield{author}{\bibinfo{person}{Andrew~L. Maas}, \bibinfo{person}{Raymond~E.
  Daly}, \bibinfo{person}{Peter~T. Pham}, \bibinfo{person}{Andrew~Y. Ng}, {and}
  \bibinfo{person}{Christopher Potts}.} \bibinfo{year}{2011}\natexlab{}.
\newblock \showarticletitle{Learning Word Vectors for Sentiment Analysis}. In
  \bibinfo{booktitle}{\emph{ACL}}.
\newblock


\bibitem[\protect\citeauthoryear{Nguyen, Gravel, Trieschnigg, and Meder}{Nguyen
  et~al\mbox{.}}{2013}]%
        {Nguyen2013HowOD}
\bibfield{author}{\bibinfo{person}{Dong Nguyen}, \bibinfo{person}{Rilana
  Gravel}, \bibinfo{person}{Dolf Trieschnigg}, {and} \bibinfo{person}{Theo
  Meder}.} \bibinfo{year}{2013}\natexlab{}.
\newblock \showarticletitle{"How Old Do You Think I Am?" A Study of Language
  and Age in Twitter}. In \bibinfo{booktitle}{\emph{ICWSM}}.
\newblock


\bibitem[\protect\citeauthoryear{O'Keeffe, Clarke-Pearson,
  et~al\mbox{.}}{O'Keeffe et~al\mbox{.}}{2011}]%
        {o2011impact}
\bibfield{author}{\bibinfo{person}{Gwenn~Schurgin O'Keeffe},
  \bibinfo{person}{Kathleen Clarke-Pearson}, {et~al\mbox{.}}}
  \bibinfo{year}{2011}\natexlab{}.
\newblock \showarticletitle{The impact of social media on children,
  adolescents, and families}.
\newblock \bibinfo{journal}{\emph{Pediatrics}} \bibinfo{volume}{127},
  \bibinfo{number}{4} (\bibinfo{year}{2011}), \bibinfo{pages}{800--804}.
\newblock


\bibitem[\protect\citeauthoryear{Prettenhofer and Stein}{Prettenhofer and
  Stein}{2011}]%
        {Prettenhofer2011CrossLingualAU}
\bibfield{author}{\bibinfo{person}{Peter Prettenhofer} {and}
  \bibinfo{person}{Benno Stein}.} \bibinfo{year}{2011}\natexlab{}.
\newblock \showarticletitle{Cross-Lingual Adaptation Using Structural
  Correspondence Learning}.
\newblock \bibinfo{journal}{\emph{ACM Trans. Intell. Syst. Technol.}}
  (\bibinfo{year}{2011}).
\newblock


\bibitem[\protect\citeauthoryear{Rao, Yarowsky, Shreevats, and Gupta}{Rao
  et~al\mbox{.}}{2010}]%
        {rao2010classifying}
\bibfield{author}{\bibinfo{person}{Delip Rao}, \bibinfo{person}{David
  Yarowsky}, \bibinfo{person}{Abhishek Shreevats}, {and}
  \bibinfo{person}{Manaswi Gupta}.} \bibinfo{year}{2010}\natexlab{}.
\newblock \showarticletitle{Classifying latent user attributes in twitter}. In
  \bibinfo{booktitle}{\emph{Proceedings of the 2nd international workshop on
  Search and mining user-generated contents}}. \bibinfo{pages}{37--44}.
\newblock


\bibitem[\protect\citeauthoryear{Sloan, Morgan, Burnap, and Williams}{Sloan
  et~al\mbox{.}}{2015}]%
        {sloan2015tweets}
\bibfield{author}{\bibinfo{person}{Luke Sloan}, \bibinfo{person}{Jeffrey
  Morgan}, \bibinfo{person}{Pete Burnap}, {and} \bibinfo{person}{Matthew
  Williams}.} \bibinfo{year}{2015}\natexlab{}.
\newblock \showarticletitle{Who tweets? Deriving the demographic
  characteristics of age, occupation and social class from Twitter user
  meta-data}.
\newblock \bibinfo{journal}{\emph{PloS one}} \bibinfo{volume}{10},
  \bibinfo{number}{3} (\bibinfo{year}{2015}).
\newblock


\bibitem[\protect\citeauthoryear{Thongtan and Phienthrakul}{Thongtan and
  Phienthrakul}{2019}]%
        {Thongtan2019SentimentCU}
\bibfield{author}{\bibinfo{person}{Tan Thongtan} {and}
  \bibinfo{person}{Tanasanee Phienthrakul}.} \bibinfo{year}{2019}\natexlab{}.
\newblock \showarticletitle{Sentiment Classification Using Document Embeddings
  Trained with Cosine Similarity}. In \bibinfo{booktitle}{\emph{ACL}}.
\newblock


\bibitem[\protect\citeauthoryear{Wang, Li, and Luo}{Wang
  et~al\mbox{.}}{2016a}]%
        {j3}
\bibfield{author}{\bibinfo{person}{Yu Wang}, \bibinfo{person}{Yuncheng Li},
  {and} \bibinfo{person}{Jiebo Luo}.} \bibinfo{year}{2016}\natexlab{a}.
\newblock \showarticletitle{Deciphering the 2016 U.S. Presidential Campaign in
  the Twitter Sphere: A Comparison of the Trumpists and Clintonists}. In
  \bibinfo{booktitle}{\emph{International Conference on Weblogs and Social
  Media}}.
\newblock


\bibitem[\protect\citeauthoryear{Wang and Luo}{Wang and Luo}{2017}]%
        {j6}
\bibfield{author}{\bibinfo{person}{Yu Wang} {and} \bibinfo{person}{Jiebo Luo}.}
  \bibinfo{year}{2017}\natexlab{}.
\newblock \showarticletitle{Gender Politics in the 2016 Presidential Election:
  A Computer Vision Approach}. In \bibinfo{booktitle}{\emph{International
  Conference on Social Computing, Behavioral-Cultural Modeling \& Prediction
  and Behavior Representation in Modeling and Simulation}}.
\newblock


\bibitem[\protect\citeauthoryear{Wang, Luo, Niemi, Li, and Hu}{Wang
  et~al\mbox{.}}{2016b}]%
        {j4}
\bibfield{author}{\bibinfo{person}{Yu Wang}, \bibinfo{person}{Jiebo Luo},
  \bibinfo{person}{Richard~G. Niemi}, \bibinfo{person}{Yuncheng Li}, {and}
  \bibinfo{person}{Tianran Hu}.} \bibinfo{year}{2016}\natexlab{b}.
\newblock \showarticletitle{Catching Fire via 'Likes': Inferring Topic
  Preferences of Trump Followers on Twitter}. In
  \bibinfo{booktitle}{\emph{International Conference on Weblogs and Social
  Media}}.
\newblock


\bibitem[\protect\citeauthoryear{Yin, Lv, Zhang, Xia, and Wu}{Yin
  et~al\mbox{.}}{2020}]%
        {yin2020covid}
\bibfield{author}{\bibinfo{person}{Fulian Yin}, \bibinfo{person}{Jiahui Lv},
  \bibinfo{person}{Xiaojian Zhang}, \bibinfo{person}{Xinyu Xia}, {and}
  \bibinfo{person}{Jianhong Wu}.} \bibinfo{year}{2020}\natexlab{}.
\newblock \showarticletitle{COVID-19 information propagation dynamics in the
  Chinese Sina-microblog}.
\newblock \bibinfo{journal}{\emph{Mathematical biosciences and engineering:
  MBE}} \bibinfo{volume}{17}, \bibinfo{number}{3} (\bibinfo{year}{2020}),
  \bibinfo{pages}{2676}.
\newblock


\bibitem[\protect\citeauthoryear{Yin, Cui, Chen, Hu, and Zhou}{Yin
  et~al\mbox{.}}{2015}]%
        {Yin2015DynamicUM}
\bibfield{author}{\bibinfo{person}{Hongzhi Yin}, \bibinfo{person}{B. Cui},
  \bibinfo{person}{Ling Chen}, \bibinfo{person}{Zhiting Hu}, {and}
  \bibinfo{person}{Xiaofang Zhou}.} \bibinfo{year}{2015}\natexlab{}.
\newblock \showarticletitle{Dynamic User Modeling in Social Media Systems}.
\newblock \bibinfo{journal}{\emph{ACM Trans. Inf. Syst.}}
  (\bibinfo{year}{2015}).
\newblock


\end{thebibliography}

\appendix
\small

\end{document}